\documentclass[draft]{agujournal2019}
\usepackage{url} %this package should fix any errors with URLs in refs.
\usepackage{lineno}
\usepackage[inline]{trackchanges} %for better track changes. finalnew option will compile document with changes incorporated.
\usepackage{mathtools}
\usepackage{soul}
\usepackage{ragged2e}
\usepackage{caption}
\usepackage{subfigure}
 \usepackage{makecell}
  \usepackage{graphicx}
 \usepackage{float}
\draftfalse

\journalname{JGR: Solid Earth}

\begin{document}

%% ------------------------------------------------------------------------ %%
%  Title
%
% (A title should be specific, informative, and brief. Use
% abbreviations only if they are defined in the abstract. Titles that
% start with general keywords then specific terms are optimized in
% searches)
%
%% ------------------------------------------------------------------------ %%

\title{Deep Neural Networks for Creating Reliable PmP Database with a Case Study in Southern California}

%% ------------------------------------------------------------------------ %%
%
%  AUTHORS AND AFFILIATIONS
%
%% ------------------------------------------------------------------------ %%

% Authors are individuals who have significantly contributed to the
% research and preparation of the article. Group authors are allowed, if
% each author in the group is separately identified in an appendix.)

% List authors by first name or initial followed by last name and
% separated by commas. Use \affil{} to number affiliations, and
% \thanks{} for author notes.
% Additional author notes should be indicated with \thanks{} (for
% example, for current addresses).

% Example: \authors{A. B. Author\affil{1}\thanks{Current address, Antartica}, B. C. Author\affil{2,3}, and D. E.
% Author\affil{3,4}\thanks{Also funded by Monsanto.}}

\authors{Wen Ding\affil{1}, Tianjue Li\affil{2,3}, Xu Yang\affil{4}, Kui Ren\affil{1}, and Ping Tong\affil{2,3,5}}

% \affiliation{1}{First Affiliation}
% \affiliation{2}{Second Affiliation}
% \affiliation{3}{Third Affiliation}
% \affiliation{4}{Fourth Affiliation}

\affiliation{1}{Department of Applied Physics and Applied Mathematics, Columbia University, New York, NY 10027, USA}
\affiliation{2}{Division of Mathematical Sciences, School of Physical and Mathematical Sciences, Nanyang Technological University, 637371, Singapore}
\affiliation{3}{Earth Observatory of Singapore, Nanyang Technological University, 639798, Singapore}
\affiliation{4}{Department of Mathematics, University of California, Santa Barbara, CA 93106, USA}
\affiliation{5}{Asian School of the Environment, Nanyang Technological University, 639798, Singapore}

%(repeat as many times as is necessary)

%% Corresponding Author:
% Corresponding author mailing address and e-mail address:

% (include name and email addresses of the corresponding author.  More
% than one corresponding author is allowed in this LaTeX file and for
% publication; but only one corresponding author is allowed in our
% editorial system.)

% Example: \correspondingauthor{First and Last Name}{email@address.edu}

\correspondingauthor{Xu Yang}{xuyang@math.ucsb.edu}

%% Keypoints, final entry on title page.

%  List up to three key points (at least one is required)
%  Key Points summarize the main points and conclusions of the article
%  Each must be 100 characters or less with no special characters or punctuation and must be complete sentences

% Example:
% \begin{keypoints}
% \item	List up to three key points (at least one is required)
% \item	Key Points summarize the main points and conclusions of the article
% \item	Each must be 100 characters or less with no special characters or punctuation and must be complete sentences
% \end{keypoints}

\begin{keypoints}
\item A deep-neural-network-based algorithm, PmPNet, is developed to automatically pick the Moho-reflected seismic waves PmP;
\item PmPNet efficiently achieves high precision and recall simultaneously to automatically identify PmP waves from a massive seismic database;
\item PmPNet creates a large PmP database with a total of 28,093 high-quality PmP picks in southern California.
\end{keypoints}

%% ------------------------------------------------------------------------ %%
%
%  ABSTRACT and PLAIN LANGUAGE SUMMARY
%
% A good Abstract will begin with a short description of the problem
% being addressed, briefly describe the new data or analyses, then
% briefly states the main conclusion(s) and how they are supported and
% uncertainties.

% The Plain Language Summary should be written for a broad audience,
% including journalists and the science-interested public, that will not have 
% a background in your field.
%
% A Plain Language Summary is required in GRL, JGR: Planets, JGR: Biogeosciences,
% JGR: Oceans, G-Cubed, Reviews of Geophysics, and JAMES.
% see http://sharingscience.agu.org/creating-plain-language-summary/)
%
%% ------------------------------------------------------------------------ %%

%% \begin{abstract} starts the second page

\begin{abstract}
Recent progresses in artificial intelligence and machine learning make it possible to automatically identify seismic phases from exponentially growing seismic data. Despite some exciting successes in automatic picking of the first P- and S-wave arrivals, auto-identification of later seismic phases such as the Moho-reflected PmP waves remains a significant challenge in matching the performance of experienced analysts. The main difficulty of machine-identifying PmP waves is that the identifiable PmP waves are rare, making the problem of identifying the PmP waves from a massive seismic database inherently unbalanced. In this work, by utilizing a high-quality PmP dataset (10,192 manual picks) in southern California, we develop PmPNet, a deep-neural-network-based algorithm to automatically identify PmP waves efficiently; by doing so, we accelerate the process of identifying the PmP waves. PmPNet applies similar techniques in the machine learning community to address the unbalancement of PmP datasets. The architecture of PmPNet is a residual neural network (ResNet)-autoencoder with additional predictor block, where encoder, decoder, and predictor are equipped with ResNet connection. We conduct systematic research with field data, concluding that PmPNet can efficiently achieve high precision and high recall simultaneously to automatically identify PmP waves from a massive seismic database. Applying the pre-trained PmPNet to the seismic database from January 1990 to December 1999 in southern California, we obtain nearly twice more PmP picks than the original PmP dataset, providing valuable data for other studies such as mapping the topography of the Moho discontinuity and imaging the lower crust structures of southern California.

\end{abstract}

\section*{Plain Language Summary}

The success of machine learning in computer sciences, medical sciences, and many other fields has accelerated the implementation and development of machine learning techniques in seismology, making it possible to automatically identify seismic phases from the exponentially growing seismic data. At present, the auto-identification of later seismic phases, such as the Moho-reflected PmP waves, remains a significant challenge in matching the performance of experienced analysts. The main difficulty lies in the rare identifiable PmP waves, which makes the identification problem inherently unbalanced. In this work, by utilizing a high-quality PmP dataset in southern California, we develop a deep-neural-network-based algorithm, PmPNet, to accelerate the process of identifying the PmP waves. We conduct systematic research with field data, and conclude that the PmPNet can efficiently achieve high precision and high recall simultaneously for automatically identifying the PmP waves from a massive seismic database. Applying the pre-trained PmPNet to the seismic database from January 1990 to December 1999, we have tripled the PmP dataset in southern California.

\section{Introduction} 
%%%%%%%%%%%%%%%%%%%%%%%%%%%%%%%%%%%%%%

\begin{figure}[!htb]
	\centering
	\includegraphics[width=0.9\linewidth]{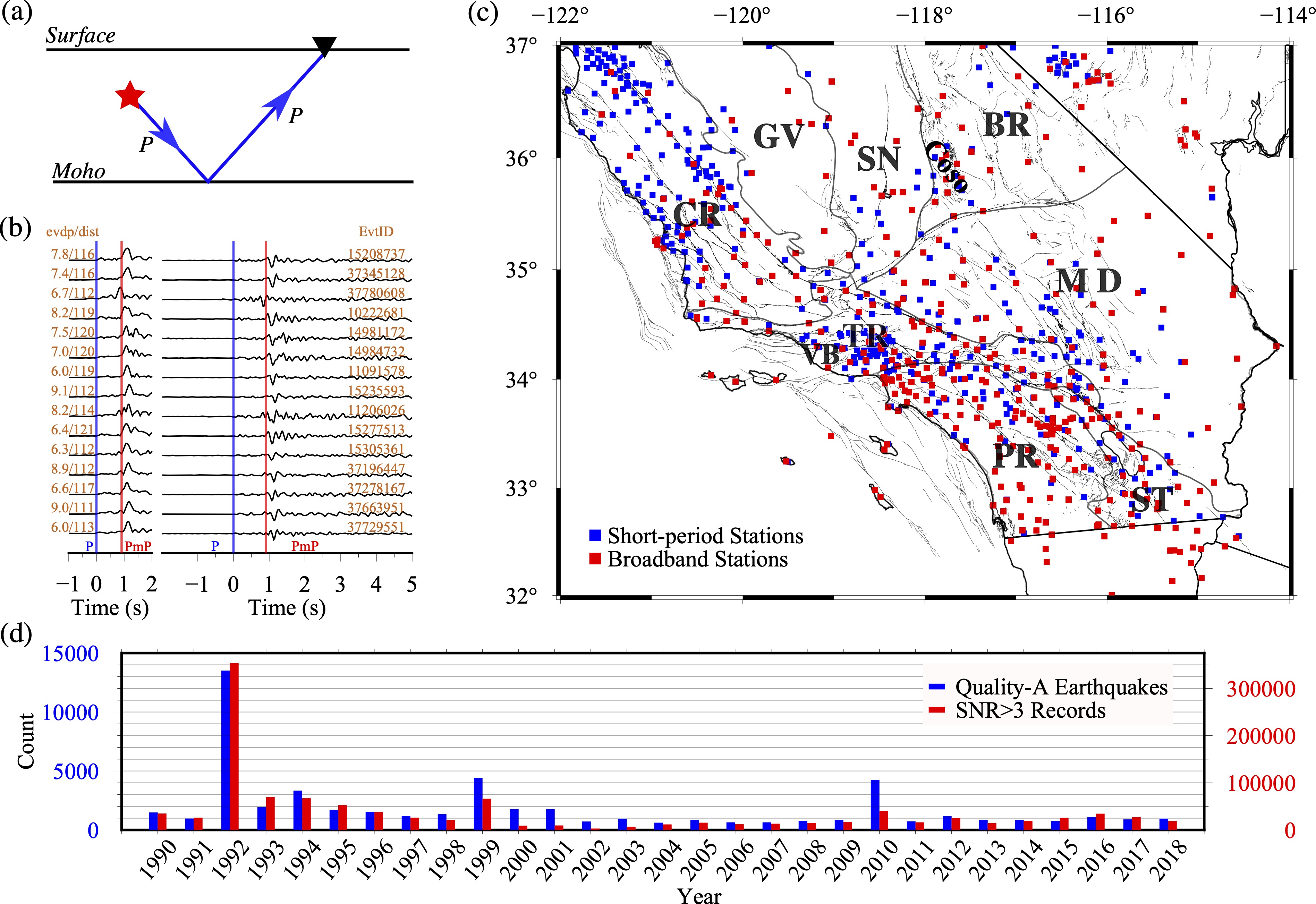}
	\caption{(a) Schematic diagram showing the ray path of PmP wave. The PmP wave (blue line) leaves the source (red star) downward, then impinges onto the Moho interface and being reflected upward, last being recorded by a surface seismic station (black triangle). (b) Typical PmP waves and their associated envelopes. These vertical-component waveforms are recorded by the broadband seismic station CI.GMR. They have been bandpass filtered (1-7Hz). The blue line denotes the oneset of the first P wave, and the red line shows the onset of PmP wave. For each waveform, its related event ID (EvtID), event focal depth (evdp) and the source-station distance (dist) are also shown at the top in orange. (c) The map shows the distribution of short-period seismic stations (blue squares) and broadband seismic stations (red squares). Main subterranes in southern California include Coast Ranges (CR), Great Valley (GV), Sierra Nevada (SN), Coso volcanic field (Coso), Basin and Range (BR), Transverse Ranges (TR), Ventura Basin (VB), Mojave Desert (MD), Peninsular Ranges (PR) and Salton Trough (ST). (d) Available seismic events each year with high-quality location, i.e., $<$ 1 km in the horizontal direction and $<$ 2 km in the vertical direction (in blue) and available seismic records each year with SNR $>$ 3 (in red).}
	\label{FIG:fig1}
\end{figure}

Continuously improving our knowledge of the crust is of fundamental importance, since it preserves a more than 3.4 Gy-old record of the planet's evolution, provides us with natural resources, and presents social challenges in the form of various natural hazards \cite{christensen1995seismic,mooney20101}. Seismic direct arrivals and those resulting from single or multiple reverberations within the crust contain important information on crustal structures and dynamics. Among them, the Moho-reflected PmP waves (Figures~\ref{FIG:fig1}a and \ref{FIG:fig1}b) have been frequently used in active-source studies to constrain the mid-lower crustal structure~\cite{mooney20101}. However, the scarcity of identifiable high-quality PmP waves in earthquake seismograms, i.e., $\sim 1\%$~\cite<Table 2 in the supporting information;>{li2022moho,wang2018crustal,sun2008seismic,xia2007mapping}, hampers its wide-ranged utilities in passive-source seismic studies, especially when one needs to prepare the data through the labor-intensive manual picking work. To date, this labor-intensive picking work is inevitable. 

The success of machine learning in computer sciences, medical sciences, and many other fields has accelerated the implementation and development of machine learning techniques in seismology. During the past few years, we have seen that tasks of earthquake monitoring, including detection, hypocenter location, phase identification, and arrival time picking, can be more efficiently performed by adopting machine learning techniques than traditional approaches~\cite{Beroza2021}. In particular, scientists have a long interest in automatic arrival time picking due to its labor-intensive and repeating characteristics. Thanks to decades of accumulation of high-quality seismic data and dedicated data labeling work by skilled analysts, there are quite a few successful machine learning approaches recently developed for picking seismic arrivals automatically. \citeA{zhu2019phasenet} proposed a deep neural network-based arrival-picking method called ``PhaseNet'', which can pick the P- and S-wave arrival times of local earthquakes. \citeA{wang_deep_2019} proposed a rich side-output residual network-based method called ``PickNet'' to pick the P- and S-wave arrivals of both local and regional earthquakes. \citeA{ross2018p} developed two separate convolutional neural networks (CNNs) with each CNN for one task, dealing with the P-wave arrival time picking and the first-motion polarity identification. Later on, they presented a deep recurrent neural network-based framework named ``PhaseLink'', which aims at assigning both the P- and S-wave picks to the earthquakes that generate them~\cite{ross_phaselink_2019}. Moreover, \citeA{Garcia2021mantle} tried to use a CNN-based method to identify the surface-reflected mantle seismic phase SS and its precursors.

Despite all the aforementioned successes of machine learning in automatic P and S phases picking and the automatic identification of global reflected/scattered seismic phases (with the frequency of 15-50 s), it is still a challenge for machines to match the performance of experienced analysts in the identification of local and regional seismic reflected waves such as PmP (with the frequency of 1-7 Hz). The main difficulty lies in the fact that the identifiable high-quality PmP waves are extremely rare in the amount of 100-10,000~\cite{xia2007mapping,sun2008seismic,wang2018crustal,li2022moho} compared to the P- and S-wave data in the amount of more than 100,000~\cite{zhu2019phasenet,wang_deep_2019,ross2018p,ross_phaselink_2019}, and the local PmP waves are much more difficult to identify in such a high-frequency band compared to the long-period global seismic phase data. Because of this, the dataset for PmP phase identification is inherently unbalanced. When one handles the unbalanced dataset, it contains very few samples belonging to the interested category, and it becomes tricky to decide how to weigh the loss function in machine learning algorithms. Therefore, the resulting lack of reliable hand-labeled data makes it hard to train standard machine learning models to pick the PmP phase automatically. This is unlike the P and S phase picking where one in general has enough well-labeled data to train regular neural networks. 

In this study, we propose PmPNet, one deep learning algorithm enhanced from the autoencoder framework~\cite{6302929} with convolutional residual connections~\cite{he2015deep}, to accelerate the process of identifying the PmP phase. Moreover, our algorithm can overcome the challenge caused by unbalanced training datasets in imaging and other applications~\cite{johnson_survey_2019}. Based on a high-quality PmP dataset (10,192 manual picks) in southern California, We demonstrate that the proposed algorithm can achieve high precision and recall for the automatic identification of PmP waves from a large number of seismograms. Applying the developed PmPNet to the more than 30-years accumulated seismic data (Figures~\ref{FIG:fig1}c and \ref{FIG:fig1}d), we update the PmP database for southern California with more than 28,000 picks. 

%%%%%%%%%%%%%%%%%%%%%%%%%%%%%%%%%%%%%%
\section{Method}
\label{SEC:Method}
%%%%%%%%%%%%%%%%%%%%%%%%%%%%%%%%%%%%%%

%%% We shall start by introducing the overall PmPNet architecture, illustrated in Figure~\ref{FIG:PmPNet}.

%%%%%%%%%%%%%%%%%%
\subsection{PmPNet architecture}
%%%%%%%%%%%%%%%%%%

%Convolutional neural networks (CNN), residual neural networks (ResNet) and autoencoders (AE) have been extremely successful in image processing applications~\cite{lecun_deep_2015,he2015deep,bank2021autoencoders}. 
In this study, we implemented PmPNet as a convolutional residual autoencoder (RAE) with an additional prediction module.
\begin{figure}[!htb]
\centering
\includegraphics[width=\textwidth]{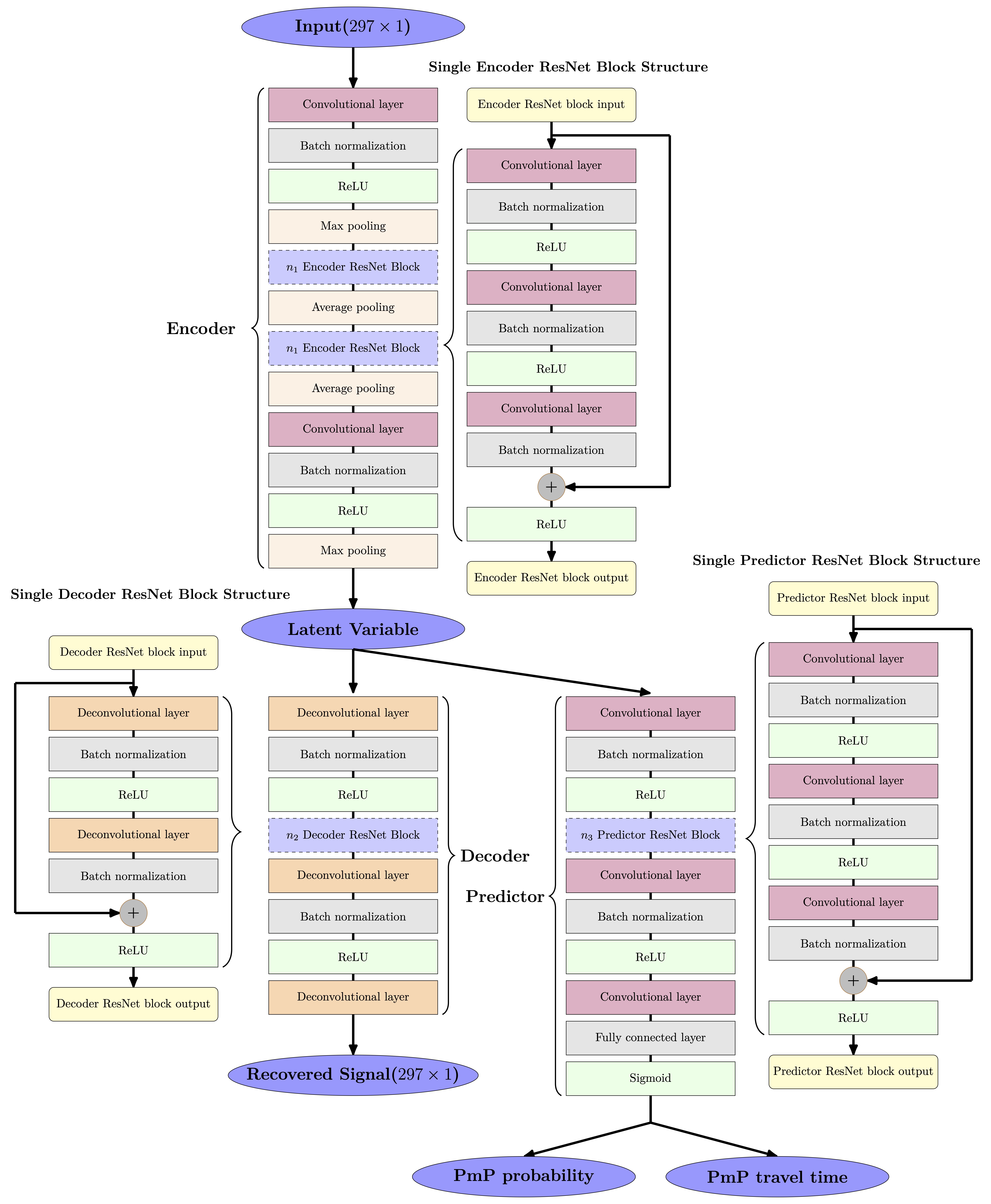}
\caption{The main architecture of the PmPNet and the data flow inside the network: The input of the PmPNet is a one-dimensional vector combining three parts: {{signal envelope, epicentral distance}}, and {{focal depth}}; PmPNet outputs three quantities: the recovered {signal (including {signal envelope, epicentral distance}, and {focal depth})}, the PmP probability $p$ and the PmP travel time $t$; PmPNet includes three major substructures: an encoder, a decoder, and a predictor.}
\label{FIG:PmPNet}
\end{figure}

%%%
\subsubsection{Standard autoencoders}
%%%

Standard AEs are trained to regenerate their inputs. Training an AE aims to make its output as close to the corresponding input as possible. To be more specific, we treat an AE as a black box, and denote the input signal as $x$ where $x$ could be a seismic signal in our application. We denote the loss function as $l(\cdot, \cdot)$. The loss function could be the usual $L_2$ loss function $l(x_1, x_2) := \|x_1-x_2\|_2^2$ or any alternative. With this notation, the loss of AE at input $x$ is $l(AE(x), x)$. In other words, AE is trained to resemble the identity operator. A more detailed structure of AE consists of two substructures: an encoder and a decoder. The encoder takes in an input $x$ and maps the input to a latent variable $z$, while the decoder takes in the latent variable and maps it to the output $AE(x)$. We write the process abstractly as $z = Encoder(x)$ and $AE(x) = Decoder(z)$.

The benefit of AEs lies in their ability to perform nonlinear feature embedding from the high dimensional input space to the low dimensional latent space. Suppose that an AE is well trained (in the sense that $x$ is indeed very close to $AE(x)$, in appropriate metric). Then $z = Encoder(x)$ and $x\approx Decoder(z)$. Therefore knowing either $x$ or $z$ is the same as knowing both. In most cases, the dimension of $z$ is far lower than $x$. Therefore $z$ is a low dimensional representation of original input $x$. Due to this simple fact that $x$ and $z$ contain the same amount of information (since we could reproduce one from the other), replacing the high-dimensional $x$ with the low-dimensional $z$ is possible and is computationally beneficial (again, since $z$ has a lower effective dimension).

%%%
\subsubsection{Overview of PmPNet structure}
%%%

In our construction, the input of the PmPNet is a one-dimensional vector of length $297$. It combines three parts: \textit{envelope, dist, evdp}. The length $281$ envelope is the normalized envelope of {the vertical component of seismic signals}, which has been resampled at $40$ Hz and covers the time window from $2$ s before to $5$ s after the observed P-wave arrival. Dist refers to the epicentral distance, and evdp refers to the focal depth, both being repeated $8$ times and concatenated to the end of the signal. The number of duplication $8$ is chosen according to our experiments, and the final results presented in this paper are not sensitive to this selection. To summarize, the input of the PmPNet is of the form:
\begin{linenomath*}
\[
input\ \ x = (\underbrace{envelope_1,\cdots, envelope_{281}, \underbrace{dist, \cdots, dist}_{8}, \underbrace{evdp, \cdots, evdp}_{8}}_{297})
\]
\end{linenomath*}
PmPNet outputs three quantities: (i) the recovered input, (ii) the PmP travel time $t$ (a positive real number), and (iii) the PmP probability $p$, a real number in $[0,1]$ representing the probability that the input seismic signal contains a PmP phase.

PmPNet includes three major substructures: an encoder, a decoder, and a predictor. Similar to a standard AE, the encoder and the decoder are trained to regenerate the input. We train the predictor to read the latent variable (generated by the encoder) to predict the PmP probability $p$ and travel time $t$. To be more precise, let $\tilde{x}$ be the output of the decoder, i.e., the recovered input. Then $z = Encoder(x)$, $\tilde{x} = Decoder(z)$, and $(p, t) = Predictor(z)$. Graphical data flows inside the PmPNet are described in Figure~\ref{FIG:PmPNet}. 

%%%
\subsubsection{Substructures of PmPNet}
%%%

Each substructure of the PmPNet is a convolutional neural network with residual connections. Here are more details on each of the substructures, as illustrated in Figure~\ref{FIG:PmPNet}:
\begin{itemize}
	\item The encoder includes several convolutional layers~\cite{LecunCNN}, batch normalization layers~\cite{ioffe2015batch}, ReLU layers~\cite{Nair2010RectifiedLU}, pooling layers~\cite{NIPS1989_53c3bce6} and $n_1$ encoder ResNet blocks~\cite{he2015deep}.
	\item The decoder consists of several de-convolutional layers, batch normalization layers, ReLU layers, pooling layers and $n_2$ decoder ResNet blocks.
	\item The predictor involves several convolutional layers, batch normalization layers, ReLU layers, pooling layers, and $n_3$ predictor ResNet blocks as well as fully connected layer to make PmP probability and travel time prediction.
\end{itemize}
The exact number of blocks $n_1$, $n_2$, and $n_3$ are determined through benchmarking the performance of the PmPNet when training with real data. We refer interested readers to~\cite{schmidhuber_deep_2015} for a quick overview of the various components of the network we used in our construction.

%%%%%%%%%%%%%%%%%%
\subsection{Dataset preparation}
\label{SUBSEC:Data Preparation}
%%%%%%%%%%%%%%%%%%

{We have developed a two-stage workflow of identifying and picking PmP waves in a semiautomatic way~\cite{li2022moho}. Briefly speaking, the two-stage workflow includes two parts: At the first stage, high-quality PmP waves are automatically picked on selected seismograms. At the same time, a visual check on three-component waveforms is conducted to confirm that the chosen signals indeed come from Moho reflection. At the second stage, the volume of the PmP dataset is expanded by involving other waves traveling along similar paths as those picked at the first stage.} By utilizing the newly developed two-stage workflow, we have built the first PmP database with 10,192 PmP waves from the broadband vertical-component seismic data retrieved from southern California Earthquake Data Center~\cite{scedc2013southern}. Here, we extract 5,000 waveforms that contain PmP waves from the PmP database and assign a PmP label to each waveform. We also find additional 100,000 waveforms that do not contain obvious PmP waves and assign a non-PmP label to each waveform. Thus our training data contain those labeled with either PmP or non-PmP with the data amount ratio of 1:20 to mimic the rare high-quality PmP waves in the real case. The used seismic data spanning from January 2000 to July 2010  are triggered by 6,636 local earthquakes with a magnitude between 2.0 and 5.0, epicentral distance (dist) between 50 and 200 km, and focal depth (evdp) shallower than 20 km. We have resampled the raw seismic waveforms to a uniform rate of 40 Hz and bandpass filtered (1-7 Hz) to prepare the network's input. We then change the velocity seismograms into envelopes, and only the portion of an envelope in a time window from 2 s before to 5 s after the first P is used. The PmP travel times for the waveforms labeled with PmP are manually picked, whereas for those labeled with non-PmP, the associated PmP travel times are calculated by using a uniform 1D P-wave velocity model~\cite<the HK model,>{hadley1977seismic} with a fixed Moho at 30 km depth. 

Before the training dataset is fed to the PmPNet, {we first normalize each sample of the input envelope to have a maximum one, and then independently standardize each time step of the input envelope to mean zero and variance one, and then linearly normalize the dist, evdp features into the range of $[0, 1]$}.

%Such normalization plays a crucial role in stabilizing the training process. For example, ...(no normalization for PmP travel time, and different scaling processes applied to the input signal) 

%%%%%%%%%%%%%%%%%%
\subsection{Dealing with the issue of unbalanced data}
%%%%%%%%%%%%%%%%%%

With the help of an expert hand-picked PmP phase dataset, we can formulate the PmP phase-detection problem as a supervised classification problem where the objective is to train a binary classifier that would allow us to divide a given set of seismic signals into a category of signals with a PmP phase and another category without it. The main challenge in the training of the classifier is that the training dataset contains a significantly larger amount of signals with no PmP phase, the majority class, than signals with a PmP phase, the minority class. This unbalance of the dataset is intrinsic since PmP waves are less abundant than the {\rm P} and {\rm S} waves or maybe equally abundant but less frequently identified by domain experts. Naive classifiers trained with such an unbalanced dataset will typically over-classify the majority class due to their excessive exposure to the majority samples. This means that the commonly-used predictive accuracy is not a sensible performance measure for a classifier trained with unbalanced data. To give an example, consider a PmP dataset with $0.1\%$ positive PmP phase signals. Any naive PmP classifier can achieve $99.9\%$ accuracy on this dataset by simply labeling all signals as non PmP. %Therefore on a dataset with $0.1\%$ positive PmP phase signals by simply labeling all signals as non PmP. Obviously, such a classifier would be meaningless. 

Developing methods to handle the challenge of the unbalanced dataset is an important research topic in the machine learning literature; see for instance~\cite{johnson_survey_2019,haixiang_learning_2017} for recent reviews on the subject. In this work, we adopt an algorithmic level method to handle the issue of unbalanced data. The method is based on precision and recall values, instead of predictive accuracy, as the evaluative metric for our PmPNet to be trained. %performance measure of the   the observation that in the case of unbalanced training data, predictive accuracy is not a sensible evaluative metric for the classifier to be trained. For instance, in our case, because the prediction is dominated by the majority class, i.e. the non PmP wave in our case. 

%The disadvantage of using predictive accuracy as the evaluative metric in the case of unbalanced data is obvious: 

%for a classifier, it works well only if there are approximately equal number of samples belonging to each class. When it comes to handle unbalance dataset, however, it offers no insight, because the prediction is dominated by the majority class, i.e. the non PmP wave in our case. As an example, a naive PmP classifier can achieve $99.9\%$ accuracy on a dataset with $0.1\%$ positive PmP phase signals by simply labeling all signals as non PmP. Obviously, such a classifier would be meaningless. 

%The method is based on the precision-recall value instead of predictive accuracy as the performance measure for the classifier to be trained.

%%%
\subsubsection{Performance measures for classification with unbalanced data}
%%%

%Precision and recall are two frequent alternative metrics for performance evaluation in the case of unbalanced data. 
To introduce the definition of precision and recall, we define the true positive ($TP$), false positive ($FP$), true negative ($TN$), and false negative ($FN$) of the PmPNet prediction from a seismic signal as follows, respectively. True positive denotes the event that the PmPNet predicts the signal has a PmP phase and the prediction is correct (that is, the signal indeed has a PmP phase). False positive denotes the event that PmPNet predicts the signal has a PmP phase, but the prediction is incorrect (that is, the signal actually has no PmP phase). True negative denotes the event that PmPNet predicts the signal has no PmP phase and the prediction is correct. False negative denotes the event that PmPNet predicts the signal has no PmP phase, but the prediction is incorrect (meaning that the signal actually has a PmP phase). We recall the following three metrics based on $TP, FP, TN, FN$:
\begin{itemize}
	\item ${\rm Precision} := \frac{TP}{TP+FP}$ measures the proportion of the PmP labeled samples by PmPNet that actually contain PmP phase. 
	\item ${\rm Recall} := \frac{TP}{TP+FN}$ measures the proportion of actual PmP phase signals that are correctly labeled to be PmP by PmPNet.
	\item {${\rm F1\  score} := 2\frac{Precision*Recall}{Precision + Recall}$ as a balanced measurement of both precision and recall.}

\end{itemize} 

A perfect PmP classifier would have a $100\%$ precision and $100\%$ recall simultaneously. However, in reality, there is a trade-off between precision and recall. If we aim at a high recall PmP phase classifier, we would want to lose as few actual PmP waves as possible, but the risk of mistaking non-PmP wave as PmP is higher, hence lowering the precision. Similarly, if the goal is to train a high precision PmP phase classifier, then we would want our labeled PmP to have a very high probability of being an actual PmP wave, i.e., the ``concentration" of PmP wave from the original dataset, but then we could be at risk of losing a larger proportion of actual PmP wave, hence lowering the recall.

%%%
\subsubsection{Loss function for classification with unbalanced data}
%%%

%measure of  where we appropriately weight the loss function for the training process to reduce ultimate prediction bias towards the majority class, that is, the class of signals 

%In this work, we adopt an algorithmic level method where 
In the training process of PmPNet, we appropriately adopt weighted binary cross-entropy loss function for classification to reduce ultimate prediction bias towards the majority class, that is, the class of signals with no PmP phase. The true label $p_{true}$ of datum $x$ is either $0$ or $1$. We force the value of $p(x)$ to be between $0$ and $1$, which represents the probability that $x$ has a PmP phase. The loss for classification purpose for that particular prediction under the weighted binary cross-entropy loss is set to be:
\begin{linenomath*}
\begin{equation}\label{EQ:Weighted Loss}
	{\rm loss_c}(p(x); p_{true}):= -\omega\, p_{true} \log(p(x))-(1-p_{true}) \log(1-p(x))
\end{equation}
\end{linenomath*}
Here $\omega > 0$ is a parameter for emphasizing precision or recall for the training process, where large $\omega$ increases the recall for {\rm PmPNet}, whereas small $\omega$ increases the precision for {\rm PmPNet}. When $\omega \gg 1$, the loss function will emphasize $-p_{true} \log(p(x))$ term, the resulting effect is that  $p(x)$ is pushed toward $1$ when $p_{true}= 1$, i.e., thus increase the recall. On the contrast, when $\omega \ll 1$, the loss function will emphasize $-(1-p_{true}) \log(1-p(x))$ term, therefore $p(x)$ is pushed down to $0$ when $p_{true}$ is $0$, hence increase precision. In our training practices in the rest of the paper, we set $\omega = 20$ to be the unbalance coefficient of the dataset to balance the importance of precision and recall. The training and validation results do not change much when we slightly change $\omega$.

%%%%%%%%%%%%%%%%%%
\subsection{Loss function for the PmPNet}
%%%%%%%%%%%%%%%%%%

We are now ready to formalize the PmPNet training process. Given a training input set $\{x_i\}_{i=1}^N$ with $N$ data points, the corresponding PmP label $\{p_{true,i}\}_{i=1}^N$ and PmP travel time $\{t_{true,i}\}_{i=1}^N$, we can optimize a PmPNet with trainable parameters set $\theta$. Three different loss functions ($l_1, l_2, l_3$) are utilized for encoder-decoder, classification and travel time training respectively. The total training loss of PmPNet is the sum of three individual losses
\begin{linenomath*}
\begin{equation}\label{EQ:PmPNet Loss}
	{\rm Loss}(\theta) := \dfrac{1}{N}\sum_{i=1}^N\Big[ l_1(\tilde{x}_i(\theta, x_i), x_i) + l_2(p_i(\theta, x_i), p_{true,i})+ l_3(t_i(\theta, x_i), t_{true,i})\Big]\,
\end{equation}
\end{linenomath*}
The variables involved are summarized as follows:
\begin{itemize}
	\item $x_i$ is the input datum and $\tilde{x}_i(\theta, x_i): = Decoder_\theta(Encoder_\theta(x_i))$ is the recovered datum by the encoder-decoder pair.
	\item $p_{true,i}$ is the true PmP label picked by experts, which is either $0$ or $1$. Here $p_{true,i}=1$ means that $x_i$ has a PmP phase, while $p_{true,i}=0$ means that $x_i$ does not have a PmP phase. 
	\item $t_{true,i}$ is the true PmP travel time, which is either manually picked for those labeled with PmP or theoretically computed by using the HK model for those labeled with non-PmP, as described in Section~\ref{SUBSEC:Data Preparation}.
	\item $(p_i, t_i) := Predictor_\theta(Encoder_\theta(x_i))$ with $p_i(\theta, x_i)$ and $t_i(\theta, x_i)$ being the PmP probability and the PmP travel time respectively.
	\item the first component of the loss function $l_1(x, \tilde{x}) := \Vert x-\tilde{x}\Vert_1$ is the $L_1$ loss between input datum and recovered datum.
	\item the second component of the loss function $l_2(p, p_{true}) :={\rm loss_c}(p; p_{true})$ is the weighted cross-entropy loss discussed above in~\eqref{EQ:Weighted Loss} with the weight $\omega$ chosen to be $20$.
	\item the third component of the loss function $l_3(t, t_{true}) := \vert t-t_{true}\vert$ is absolute difference between the true travel time and predicted travel time.
\end{itemize}
We emphasize that $\tilde{x}_i$, $p_i$, and $t_i$ are outputs of PmPNet, and therefore depend on the input datum $x_i$ as well as the network parameters $\theta$.

%%%%%%%%%%%%%%%%%%
\subsection{Training and validation}
%%%%%%%%%%%%%%%%%%

Besides data preparation, the most critical step in the learning approach we proposed is the training of the PmPNet. For the sake of reproducibility, we provide here the hyperparameters we used in the training process.
\begin{itemize}
	\item Learning rate: we initialize it as $0.0001$ with exponential decay of factor $2$ for every $10$ epochs.
	\item Number of ResNet blocks for encoder, decoder and predictor are set respectively as $n_1 = 2$, $n_2 = 4$ and $n_3 = 1$.
	%\item Number of \textit{Input, output} channels and kernel sizes of the convolutional and deconvolutional layers are set to be \textcolor{red}{XXX} and \textcolor{red}{YYY} respectively.
	\item Formation of the training set, i.e. number of training data points and PmP/non-PmP ratio: we use $5,000$ PmP-labeled waveforms and $100,000$ non-PmP labeled waveforms.
	\item Batch size (i.e. the number of training data points utilized in one iteration) is set as $200$ in our experiments.
	\item Number of epochs (i.e. the number of rounds of passing entire training dataset into PmPNet) is set to be $80$ in our experiments.
	\item Training optimization algorithm: we use the Adam stochastic optimization algorithm~\cite{kingma_adam_2017}.
\end{itemize} 
Once all these hyperparameters are chosen, PmPNet is fully determined by trainable parameters $\theta$. Trainable parameters in batch normalization layers are initialized with mean $0$ and variance $1$. Trainable parameters in all other layers are initialized with Gaussian distribution with means $0$ and variances depending on the width of the layers. After we initialize all layers, training PmPNet is essentially searching $\theta$ that minimize the loss function~\eqref{EQ:PmPNet Loss} for PmPNet. 

\begin{figure}[!htb]
\centering
\includegraphics[width=\textwidth]{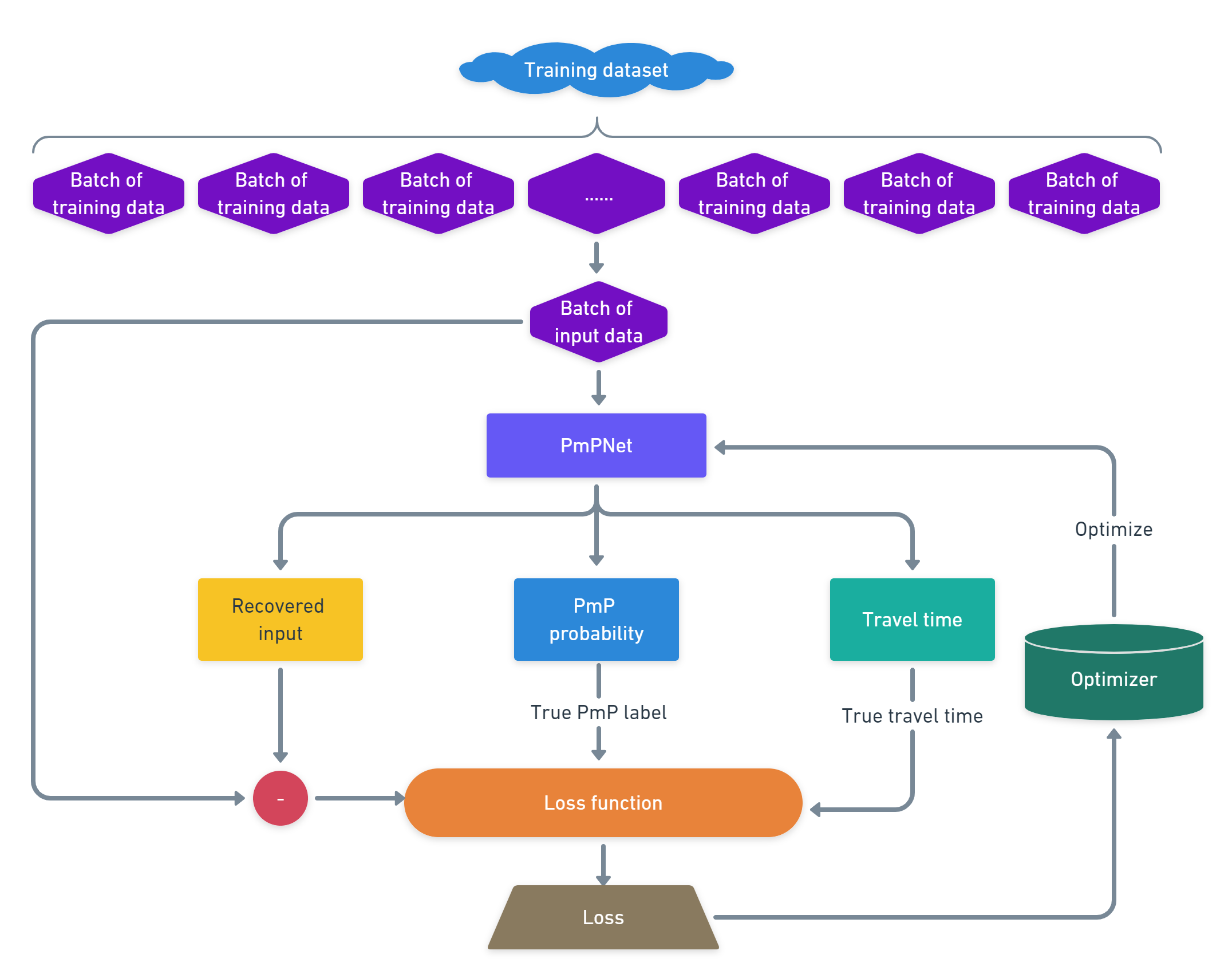}
\caption{PmPNet training flow: (i) one batch of data points is fed into PmPNet, and the loss between the PmPNet output and the true labels is computed; and (ii) the optimizer reads in the loss and update the trainable parameters of PmPNet. One epoch of training consists of continuing this iteration until the whole dataset has been tranversed. The training phase for PmPNet is complete when the pre-selected maximum number of epochs is reached.}
\label{FIG:training process}
\end{figure}
To start the training process, we randomly choose a batch of data points without replacement, and feed the data into PmPNet. PmPNet will output the corresponding reproduced signal as well as the $(p, t)$ pair with the current trainable parameter $\theta$. After computing the loss, we apply one Adam step to update the trainable parameters $\theta$. We then randomly choose another batch of data points without replacement to move another training step forward. This process continues until the whole training dataset is traversed. The training process described is often called one ``epoch". Next, we adjust the learning rate to repeat the process for another epoch. The training process is complete when the predetermined number of maximum epochs are reached. The training process for PmPNet is illustrated in Figure~\ref{FIG:training process}.

%After we initialize all layers, training PmPNet is essentially searching $\theta$ that minimize the loss function.
%\begin{eqnarray}
	%\min\limits_{\theta}{\rm Loss}(\theta) = \min\limits_{\theta}\dfrac{1}{n}\sum_{i=1}^n l_1(\tilde{x}_i(\theta, x_i), x_i) + l_2(p_i(\theta, x_i), p_{true,i})+ l_3(t_i(\theta, x_i), t_{true,i})
%\end{eqnarray}

Following common practices in the machine learning community, we perform a training-validation cycle on our PmPNet. Before the training process starts, we randomly split the input dataset (described above in Section~\ref{SUBSEC:Data Preparation}) into a training set (which is $80\%$ of the original dataset) and a validation set (which is $20\%$ of the original dataset). The training set and the validation set have no intersection, meaning that no data points in the validation set are present in the entire training process. During the validation process, we input every data point in the validation set into the post-trained PmPNet, and it outputs the corresponding $t, p$ and recovered input. We visually compare the recovered input and the input and compute the absolute difference between predicted and true travel time. In order to make classification, we set a probability threshold $p_{\rm threshold}$, that is, any signal with $p \ge p_{\rm threshold}$ is predicted as PmP wave by the {\rm PmPNet}. Otherwise, it is predicted as a non-PmP wave. Given one probability threshold, we can compute one pair of precision and recall.

%%%%%%%%%%%%%%%%%%
\subsection{PmPNet performance validation}
%%%%%%%%%%%%%%%%%%

\begin{table}[!htb]
\centering
\caption{Precision-recall pairs from PmPNet with different probability threshold for the validation dataset.}
\begin{tabular}{*{10}{c}}
\hline
$p_{threshold}$ & 0.0001& 0.01& 0.1& 0.2& 0.5& 0.8&0.9&0.99&0.999 \\
\hline
Precision & 0.4111 & 0.8387& 0.9190   & 0.9259 & 0.9464   & 0.9659 & 0.9783   & 0.9845 &  0.9865 \\
\hline
Recall & 0.9322  & 0.8908 & 0.8757    & 0.8701  & 0.8644    & 0.8531  & 0.8493    & 0.8399  &  0.8267  \\
\hline
{F1 score} & {0.5706} &   {0.8640}   & {0.8968}   & {0.8971}  &  {0.9035}  &  {0.9060}  &  {0.9092} & {0.9065}  &  {0.8996}\\
\hline
\end{tabular}
\label{tab:pr}
\end{table}
To systematically characterize the performance of the PmPNet, we construct the precision-recall curve for the validation dataset. To do that, we evaluate the precision-recall value corresponding to each classification threshold probability $p_{\rm threshold}$. We then vary the probability threshold to get the precision-recall value as a function of the threshold $p_{\rm threshold}$. The curve of precision versus recall is the precision-recall curve. This curve can help us choose the optimal value of $p_{\rm threshold}$ for future use.

\begin{figure}[!htb]
	\centering
	\includegraphics[width=0.9\linewidth]{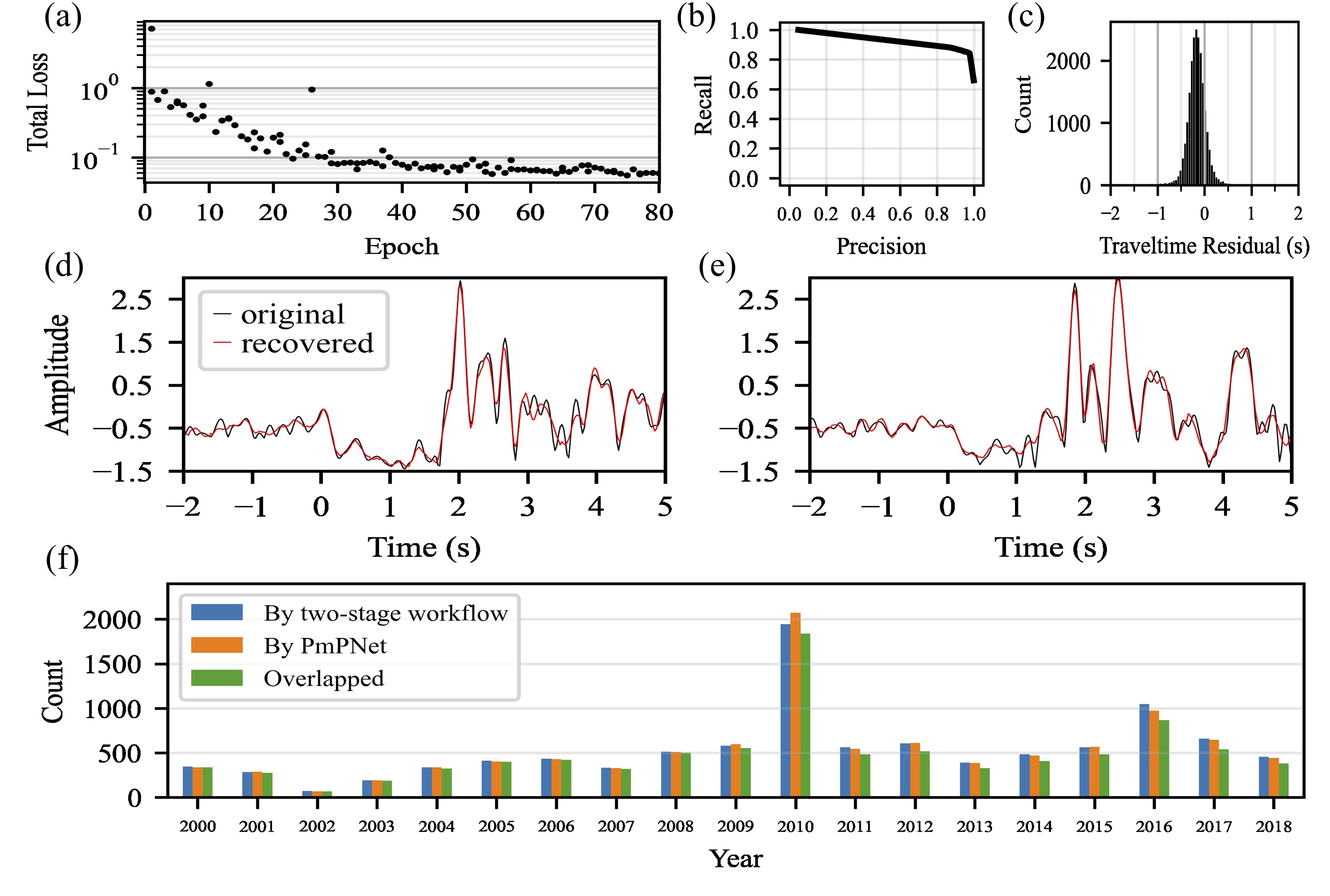}
	\caption{Training and validation performance of PmPNet when applied to real data. (a) The total training loss decreases as the epoch increases and it becomes stable after about 40-epoch training. (b) The precision-recall curve on validation set. (c)  The PmP travel time residual between the predicted and manually picked ones on validation set.  (d-e) The PmPNet recovered input and the input component on validation set. (f) PmP picks when applying the trained PmPNet to real data. Blue bars show the picked PmP waves each year by the two-stage workflow, orange bars show the identified PmP waves each year by the PmPNet with the probability of greater than 0.8, and green bars show the overlapped PmP waves each year between the two identifiers.}
	\label{FIG:fig4}
\end{figure}

The validation performances of PmPNet are as follows. The proposed PmPNet can reach high precision($96.6\%$) and recall($85.3\%$)  simultaneously, {hence a high F1 score on validation set}, see Table~\ref{tab:pr} for detailed precision-recall pairs and {F1 score} with different probability threshold and Figure~\ref{FIG:fig4}(b) for the corresponding precision-recall curve. The average travel time absolute difference is around $0.33s$, while maximum difference constantly stays within $5s$ (Figure~\ref{FIG:fig4}c). The recovered input can capture most of the patterns from the input signal (Figures~\ref{FIG:fig4}d and \ref{FIG:fig4}e), which indicates the latent variable is indeed a good representation of the input. Details of the validation performance of PmPNet are listed in Figure~\ref{FIG:fig4}. 

%%%%%%%%%%%%%%%%%%%%%%%%%%%%%%%%%%%%%%
\section{Real applications}
%%%%%%%%%%%%%%%%%%%%%%%%%%%%%%%%%%%%%%

\begin{figure}[!htb]
	\centering
	\includegraphics[width=0.9\linewidth]{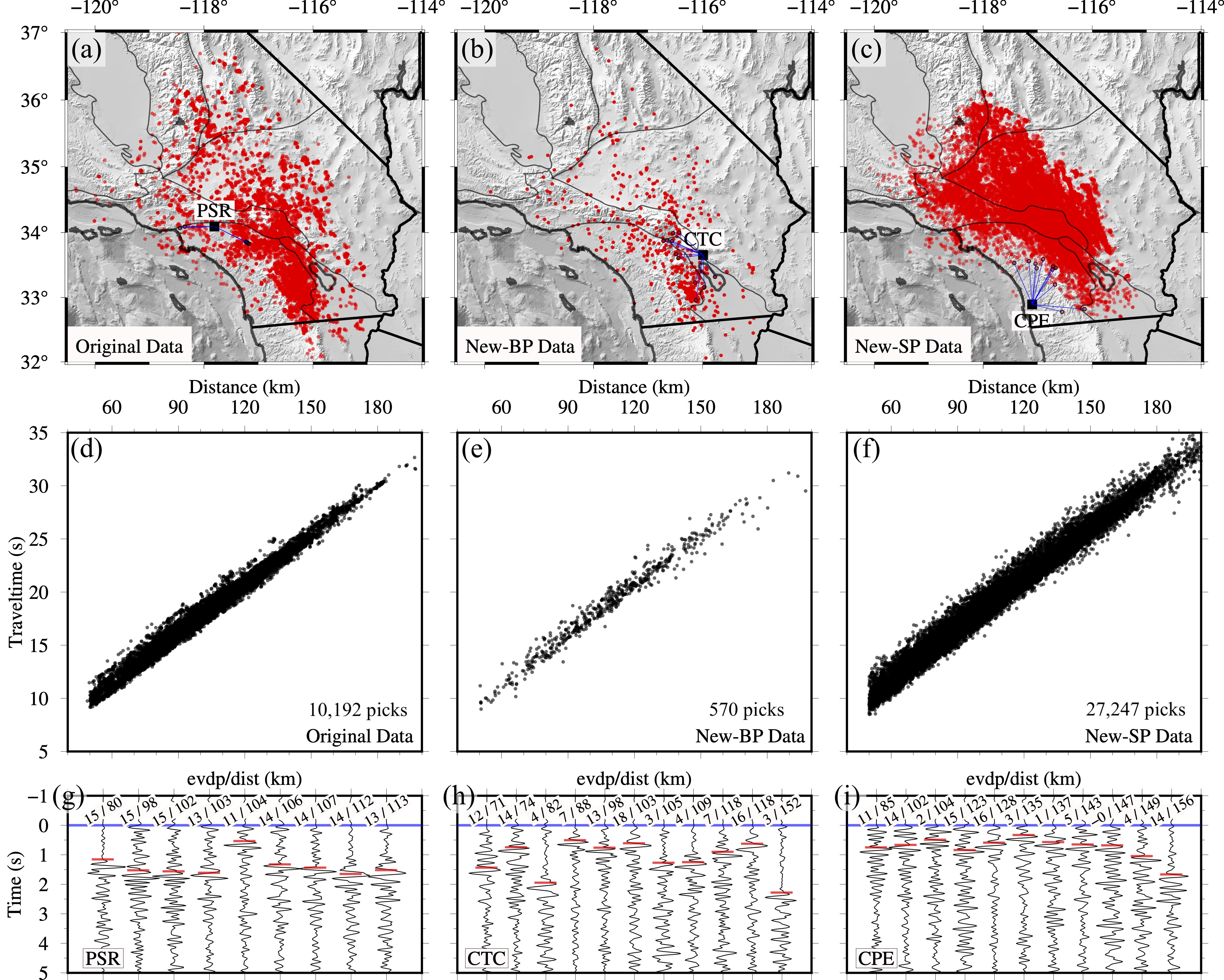}
	\caption{Comparison of the newly picked PmP waves with those built using the two-stage PmP-picking workflow. (a) The spatial distribution of PmP waves {of the original database by \protect\cite{li2022moho}}. Red points refer to the PmP refection points which are calculated in the HK model. {The black square and black circles denote the seismic station and reflection points of the PmP waveforms shown in (g). (b) Same as (a) but for the newly picked PmP waves by PmPNet on broadband (BP) data. (c) Same as (a) but for the newly picked PmP waves by PmPNet on short-period (SP) data.} (d-f) The travel times of PmP waves with respect to epicentral distance. (g-i) {Waveforms are aligned to the first P-wave arrivals and sorted according to earthquake depth (evdp) and epicentral distance (dist).} Red bars denote the onsets of PmP waves.}
	\label{FIG:fig5}
\end{figure}

For the PmP database spanning from January 2000 to December 2018 built by the two-stage PmP-picking workflow~\cite{li2022moho}, we observe that travel times of the PmP waves increase linearly with respect to the epicentral distance, amplitudes of the PmP waves are mostly larger than the P waves and particle motions of most PmP waves polarize in a similar way as the P waves.  Here, we apply the trained PmPNet to the same 19-year long vertical-component seismic data to automatically identify the waveforms which could contain high-quality PmP waves. To achieve the goal, we first select the waveforms with the PmP label with a probability of larger than 0.8. Then, we directly remove the overlapped waveforms from the selected waveforms with those picked already through the two-stage PmP-picking workflow. After that, we visually check and only keep the retained waveforms when main crustal phases like PmP, S and SmS can be observed simultaneously on the associated three-component waveforms. As shown in Figure~\ref{FIG:fig4}f, the trained PmPNet has successfully recalled the most PmP waves (larger than 96 \%) before 2011, and even for the seismic data after 2010 which are not involved in training the PmPNet, there is also a high recall value of larger than 85 \%. In addition, we have newly picked 570 high-quality PmP waves based on those picks recognized by the PmPNet (e.g., Figure~\ref{FIG:fig5}h), which have a similar data coverage (Figure~\ref{FIG:fig5}b) and a linear relationship between the PmP travel time and epicentral distance (Figure~\ref{FIG:fig5}e). All of these show the high performance of our trained PmPNet in identifying the PmP waves from broadband seismic data. 

To further release the power of our trained PmPNet in identifying the PmP waves from the massive seismic data, we also apply the trained network to the short-period vertical-component seismic data spanning from January 1990 to December 1999 retrieved also from SCEDC. The retrieved seismic waveforms are triggered by earthquakes with a magnitude between 2.0 and 5.0, and they are first cleared with the vertical-component SNR larger than 3.0, focal depth shallower than 20 km and epicentral distance between 50 and 200 km. Then, we align waveforms to the first P arrivals using the STA/LTA method~\cite{trnkoczy2009understanding} and cut them into a time window from 2 s before to 5 s after the first P arrivals. The associated envelopes are fed to the trained PmPNet, and through the network's automatic judgment, we obtain the probability of the PmP label and the PmP travel time in each seismic waveform. Waveforms that have the PmP probability of larger than 0.8 are selected out. And then we visually check the waveforms and pick the point with an abrupt amplitude jump near the predicted PmP arrival time as the PmP onset. After finishing the above process, we obtain a total of more than 27,000 PmP picks. For the obtained short-period PmP picks, their travel times increase linearly with respect to the epicentral distance (Figure~\ref{FIG:fig5}f), and they share similar waveform characters (Figure~\ref{FIG:fig5}i) to those PmP waves picked from broadband data. In addition, the newly obtained short-period PmP picks improve the data coverage significantly in the central-eastern Transverse Ranges and the Mojave Desert. 

%%%%%%%%%%%%%%%%%%%%%%%%%%%%%%%%%%%%%%
\section{PmP database for southern California}
%%%%%%%%%%%%%%%%%%%%%%%%%%%%%%%%%%%%%%

\begin{figure}[!htb]
	\centering
	\includegraphics[width=0.9\linewidth]{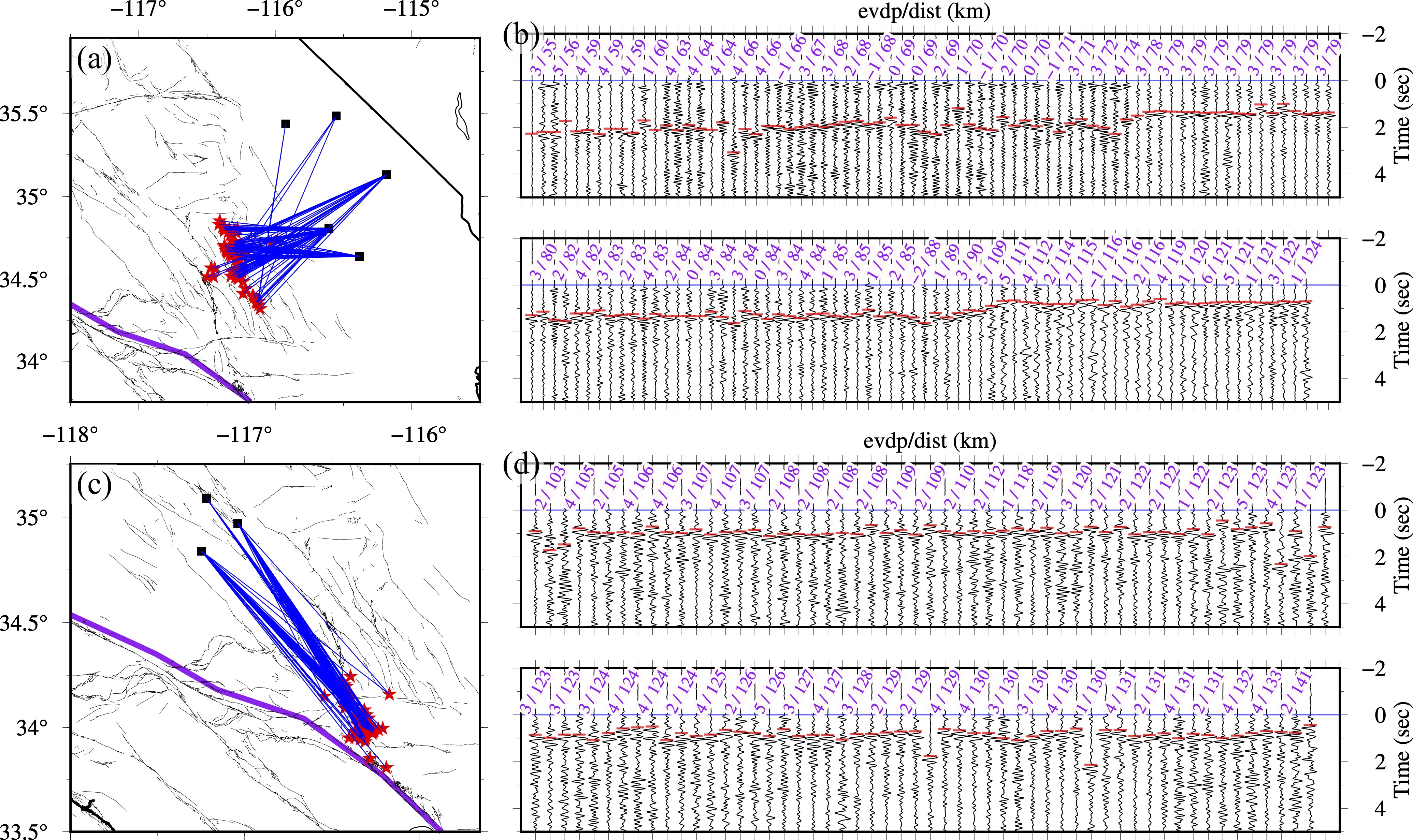}
	\caption{Using {common} reflection gather and {grouped PmP waveforms travelling along similar paths} to build the consistent PmP database. (a) An example of {common} PmP reflection gather in map view. Seismic stations are denoted as black-filled squares, and local earthquakes are represented as red stars. Red-filled circles show the reflection points of PmP waves at the Moho interface, which are calculated based on the HK model. The bold purple line shows the plate boundary between the Pacific Plate and the North American Plate. (b) The PmP waveforms within a {common} reflection gather. Red bars show the PmP arrival times. {Waveforms are aligned to the first P-wave arrivals and sorted according to earthquake depth (evdp) and epicentral distance (dist).} (c) {Same as (a) but for} an example of the PmP  {waveforms travelling along similar paths} in map view. (d) {Same as (b) but for} the PmP {waveforms travelling along similar paths}.}
	\label{FIG:fig6}
\end{figure}
\begin{figure}[!htb]
	\centering
	\includegraphics[width=0.9\linewidth]{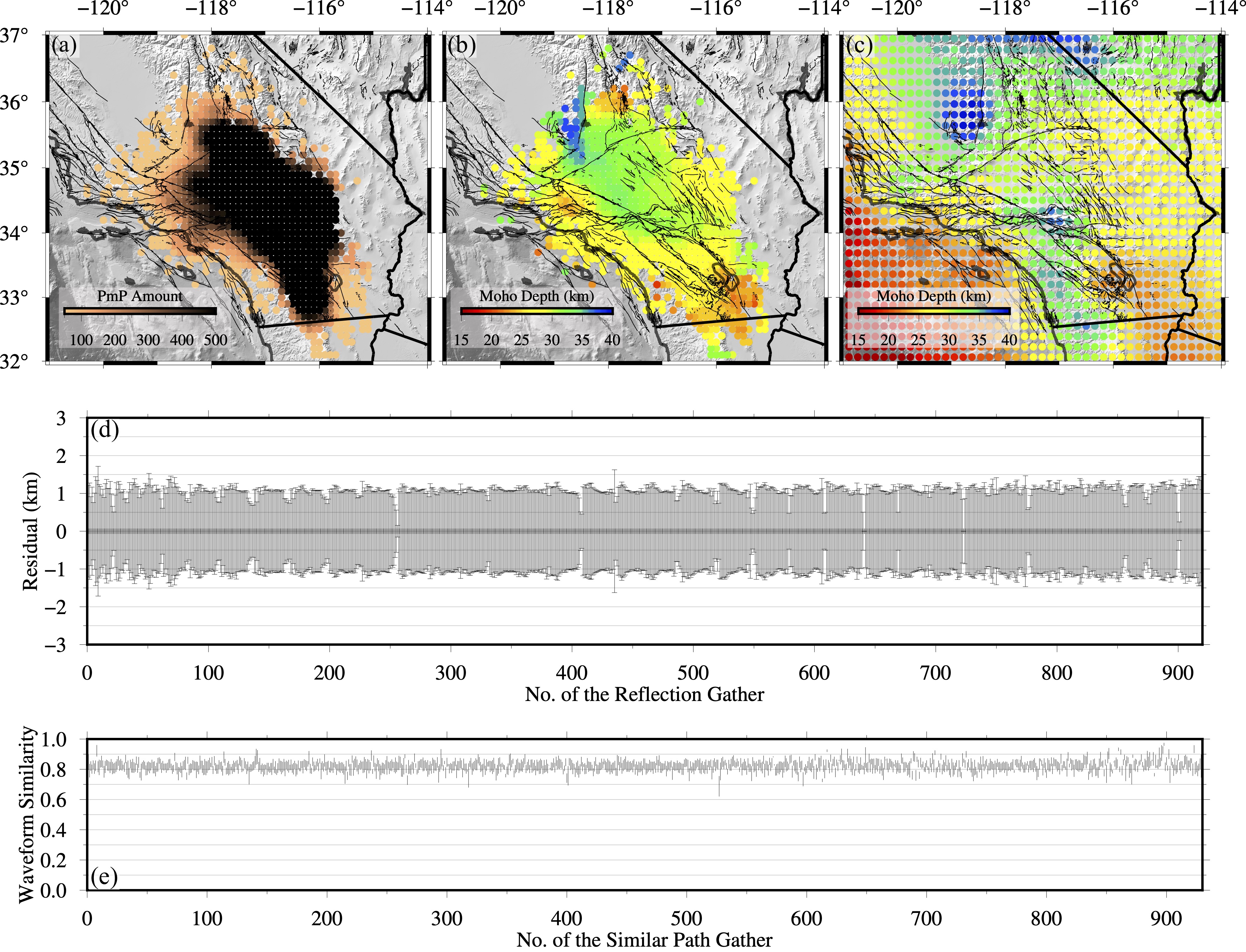}
	\caption{The newly updated PmP database for southern California. (a) {The number of PmP reflection points counted at each node of a 30 km by 30 km grid}. (b) The Moho geometry estimated from the new PmP database {at every node of the same grid in (a)}. The Moho depth is the average over a reflection gather, assuming a constant Vp = 6.3 km/s. {(c) The Moho geometry compiled in CMM-1.0, which is built mainly based on teleseismic
Ps data and active-source surveys \protect\cite{tape2012estimating}.} ({d}) The standard deviation of Moho depth for each reflection gather. ({e}) The waveform similarity (cross-correlation coefficient) for each similar ray-path gather.}
	\label{FIG:fig7}
\end{figure}

To update the high-quality PmP database in southern California, we take additional quality control on the obtained PmP waves. We first group all PmP waveforms obtained so far according to their Moho reflection points into {common} reflection gathers (Figures~\ref{FIG:fig6}a and \ref{FIG:fig6}b), where the Moho reflection points of PmP waves are calculated in the HK model and to form a gather the distance between any two reflection points is no more than 30 km. Then within each gather, we calculate the Moho depth assuming a constant ${V_p}$ = 6.3 km/s{~\cite{zhu2000moho}} for every PmP wave and discard the outliers which deviate largely in Moho depth prediction ($>$ 3.0 km) from the other co-grouped picks. After that, we retain a total of more than 28,000 high-quality PmP picks. We also group all retained PmP waveforms according to their propagation paths (Figures~\ref{FIG:fig6}c and \ref{FIG:fig6}d), and within the same {group}, the distance between any two sources of PmP waves is no more than 30 km in the horizontal direction and 3 km in the vertical direction, and the distance between any two receivers is no more than 30 km. Then within each {group}, we calculate the similarity between any two waveforms in a time window from 2 s before to 5 s after the first P arrival using a time-domain cross-correlation technique~\cite{schaff2005waveform}, and we use the averaged value as the representative waveform similarity in that {group}. The narrow Moho depth prediction range for each {common} reflection gather (Figure~\ref{FIG:fig7}c) and high representative waveform similarity for each {group} (Figure~\ref{FIG:fig7}d) show that the newly updated PmP database are of high quality and self-consistent. More field waveforms of the updated PmP database in the form of receiver gather (recorded at the same seismic station) are presented in Figures~\ref{FIG:si_fig1}-\ref{FIG:si_fig4} {in the supporting information}. As shown in Figure~\ref{FIG:fig7}a, the retained PmP waves have dense data coverage in the eastern Peninsular Ranges, the central-eastern Transverse Ranges, the Mojave Desert, and the southern part of Basin and Range. 

%%%%%%%%%%%%%%%%%%%%%%%%%%%%%%%%%%%%%%
\section{Discussion}
%%%%%%%%%%%%%%%%%%%%%%%%%%%%%%%%%%%%%%

\begin{figure}[!htb]
	\centering
	\includegraphics[width=0.9\linewidth]{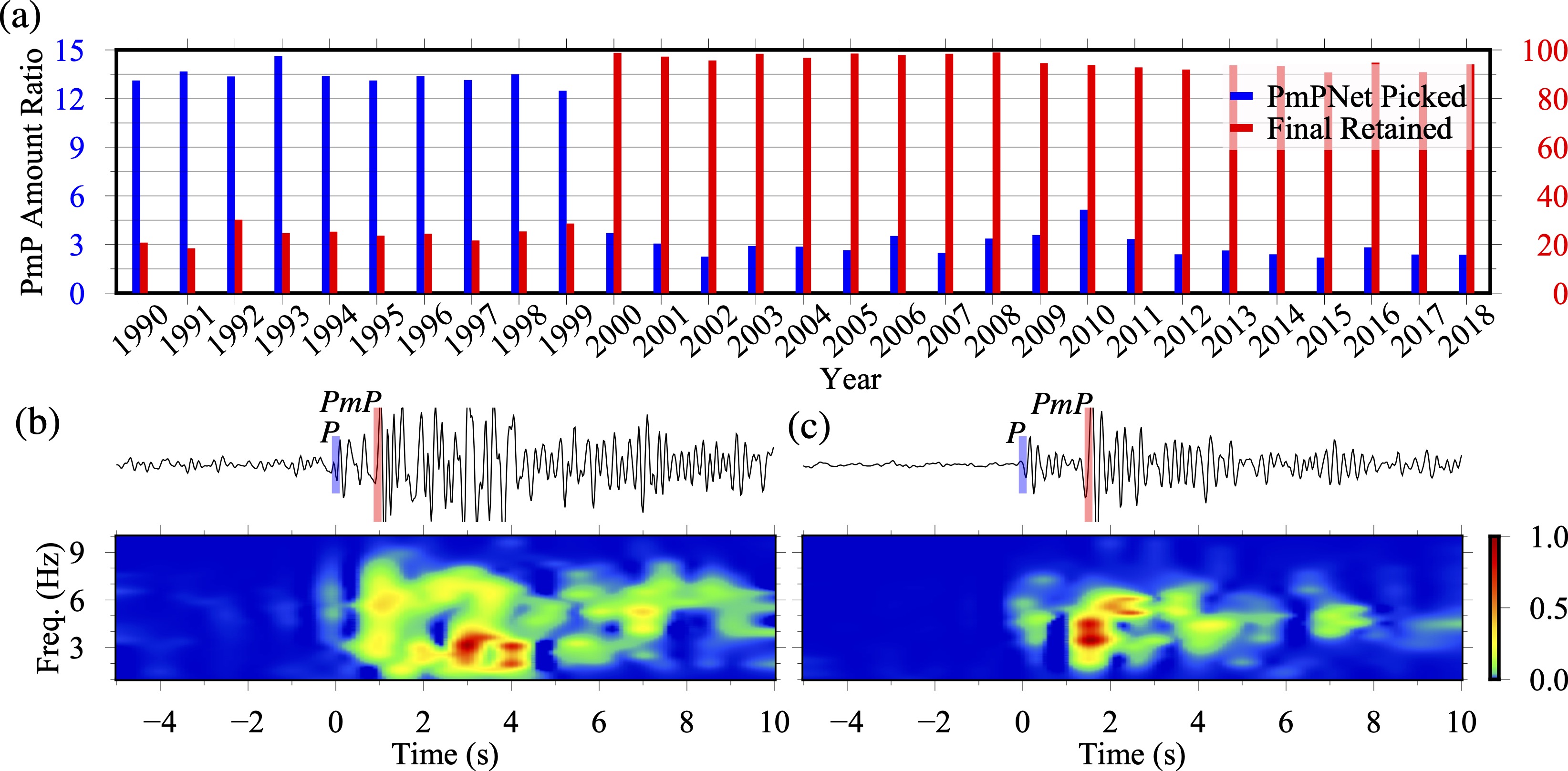}
	\caption{(a) The data amount ratio between the waveforms containing high-quality PmP with the probability of larger than 0.8 and the total available seismic data each year (in blue). And the data amount ratio between the finally retained PmP waves and the waveforms containing high-quality PmP with the probability of larger than 0.8 each year (in red). (b) An example of short-period waveform and its spectrum. Here, the shown waveform has been bandpass filtered (1-7Hz). The waveform is recorded at station CI.LJB and it is triggered by an earthquake with a magnitude of 2.0, focal depth of 3 km, and epicentral distance of 112 km. The blue bar indicates the onset time of the first P wave, and the red bar indicates the onset time of  PmP wave. (c) An example of a broadband waveform and its spectrum. The waveform is recorded at station CI.HOL and it is triggered by an earthquake with a magnitude of 2.1, focal depth of 5 km, and epicentral distance of 146 km.}
	\label{FIG:fig8}
\end{figure}
Acquiring the optimal parameters for a PmPNet with good performance requires enormous computational power, since various experiments are needed to conduct. In our experiments (Figure~\ref{FIG:si_fig5} in the supporting information), we only vary one hyperparameter with others fixed when assessing the parameter's effect on the performance of PmPNet. We found that though the number of blocks $n_1 = 2, n_2 = 4, n_3 = 1$ is simple, it results in a PmPNet performing well similar to the more complicated blocks. Using more than 2,000 waveforms labeled with PmP can generate a PmPNet performing well, whereas involving more waveforms labeled with PmP will result in a PmPNet with better performance. When using 5,000 PmP-labeled waveforms to train the PmPNet, the amount ratio between the waveforms labeled with PmP and those labeled with non-PmP {has little influence on the performance of} PmPNet. Small initialized learning rates ($< 0.001$) play a vitally important role in training a PmPNet with good performance.  The various batch size can be chosen in training a good-performance PmPNet. Using the above chosen optimal hyperparameter configuration, it takes $\sim 1.2$ hours to finish the 80-epochs training on one NVIDIA GeForce RTX 2080 graphics processing units (GPUs), and in most cases, the training loss stabilizes at around $40$th epoch (Figure~\ref{FIG:fig4}a).

It is worth mentioning that the trained PmPNet performs less well on the short-period seismic data than on the broadband seismic data. For example, the PmPNet identifies a higher percentage of waveforms on the short-period seismic data that most likely contain high-quality PmP waves, e.g., about $13.4 \%$, while we find that only about $24.3 \%$ are good PmP candidates through our manual check. Whereas on the broadband seismic data, the PmPNet identifies about $3.0 \%$ of the total available waveforms that most likely contain high-quality PmP waves (Figure~\ref{FIG:fig8}a), and through our manual check, we find that almost $95.4 \%$ are good PmP-wave candidates. The relatively poor performance of the PmPNet on the short-period seismic data is likely caused by the higher noise level compared to that on the broadband seismic data both in original waveforms (Figures~\ref{FIG:si_fig6}a and \ref{FIG:si_fig6}b {in the supporting information}) and in the bandpass filtered waveforms (1-7 Hz, Figures~\ref{FIG:fig8}b and \ref{FIG:fig8}c). Our training data (broadband seismic data) may less characterize the PmP wave on the short-period data. Involving some waveforms, either labeled PmP or non-PmP, from the short-period data in the future study could improve the performance of PmPNet in identifying the PmP waves on the short-period seismic data. 

\begin{figure}[!htb]
	\centering
	\includegraphics[width=0.9\linewidth]{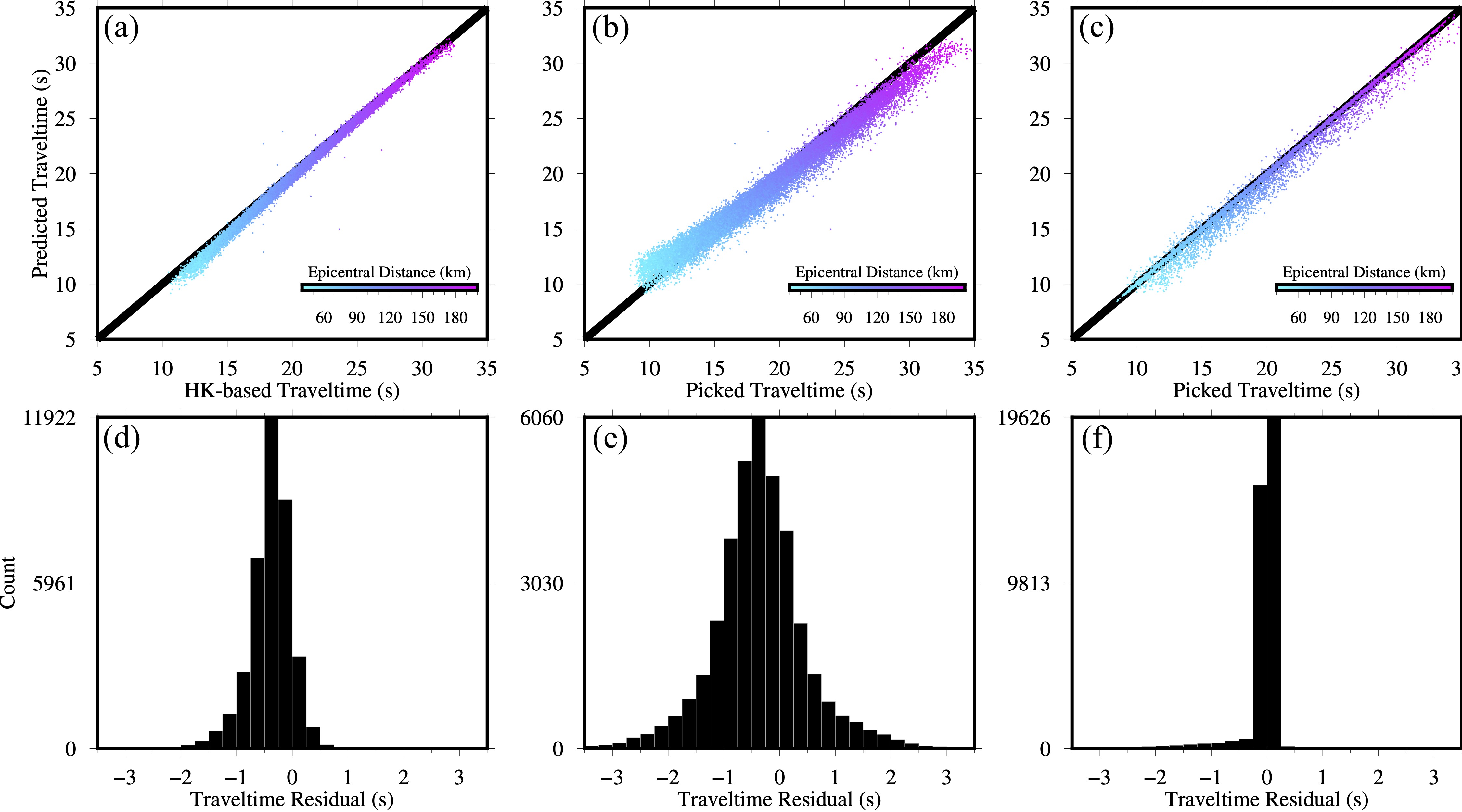}
	\caption{{Accuracy of predicted PmP traveltimes by PmPNet. (a) Comparison between the predicted PmP traveltimes by PmPNet (y-axis) and those calculated using the HK model with a fixed Moho at 30 km depth (x-axis). All manually picked data are used to carry out the test and points are color-coded with epicentral distance. (b) Same as (a) but for comparison between the predicted PmP traveltimes by PmPNet and those picked manually. (c) Same as (a) but for comparison between the predicted PmP traveltime by PmP-traveltime-Net and those picked manually. (d-f) Distribution of PmP traveltime residuals between the two datasets shown in (a-c), respectively. Traveltime residual is calculated by subtracting the x-axis data from the y-axis data.}}
	\label{FIG:fig9}
\end{figure}

{The main task for PmPNet is to identify tens of thousands of PmP waves from millions of raw seismograms. In the meantime, PmPNet also gives the absolute traveltime prediction for PmP waves. The traveltime prediction performance of PmPNet is highly comparable with the HK model (Figures~\ref{FIG:fig4}c, \ref{FIG:fig9}a and \ref{FIG:fig9}d), except for slight bias when predicting the PmP traveltimes at short distances. While we do observe that the traveltime residual between the predicted PmP traveltime by PmPNet and that picked manually is somewhat large (Figures~\ref{FIG:fig9}b and \ref{FIG:fig9}e): there are only 47$\%$ of picks with traveltime residuals no larger than 0.5 s. To solve this problem, we further train a separate model called “PmP-traveltime-Net” to predict the differential traveltime between PmP and first P waves. The structure of PmP-traveltime-Net is very similar to PmPNet, detailed introduction can be found in Text S1 in the supporting information. We train PmP-traveltime-Net on 10,000 waveform data with clear PmP waves prepared by the two-stage workflow \cite{li2022moho} and validate it on all the manually picked data (during the period from 1990 to 2018). Test performance of PmP-traveltime-Net shows that there are 95$\%$ of picks with traveltime residuals no larger than 0.5 s and even 88$\%$ of picks with traveltime residuals no larger than 0.1 s (Figures~\ref{FIG:fig9}c, \ref{FIG:fig9}f and \ref{FIG:si_fig7} in the supporting information). We recommend conducting the necessary manual verification for further research purposes, though the traveltime prediction can reach a high-level accuracy.}

For the Moho geometry constrained by the newly built PmP database (Figure~\ref{FIG:fig7}b), we observe a shallow Moho beneath the Salton Trough, and a deep Moho beneath the eastern Transverse Ranges and the southern Sierra Nevada, which are consistent with the {California} Moho {Model} 1.0~\cite<CMM-1.0,>{tape2012estimating}. However, we find the Moho beneath the Mojave Desert is slightly deeper compared to the CMM-1.0, which reflects a lower-velocity anomaly in the lower crust beneath this region~\cite{shaw2015unified,lee2014full}. In other regions like the Ventura Basin and the Coso volcanic field, we observe some small-scale Moho undulations, which is related to the complicated local crust structure \cite{yan2007regional,wilson2003single}. {In the western Peninsular Ranges, the Moho depth estimated by sparse PmP data is much shallower compared to the CMM-1.0, which may be related to the oversimplified one-layer crust model assumed here or/and a gradual transition from the crust to the upper mantle there \cite{li2022moho}.}

%%%%%%%%%%%%%%%%%%%%%%%%%%%%%%%%%%%%%%
\section{{Conclusions}}
%%%%%%%%%%%%%%%%%%%%%%%%%%%%%%%%%%%%%%

{In this work, we have proposed a deep-neural-network-based algorithm, PmPNet, to help automatically identify PmP waves. Taking advantage of the manually prepared PmP database (10,192 picks during the period 2000.01 $\sim$ 2018.12) by \citeA{li2022moho}, we use 5,000 signal envelopes with clear PmP waves (during the period 2000.01 $\sim$ 2010.07) and 100,000 signal envelopes without clear PmP waves to train (80$\%$ of envelopes) and validate (20$\%$ of envelopes) the PmPNet. We find the optimal model parameters of PmP-Net through checking three metrics: precision, recall, and F1 score. Our trained optimal PmPNet can reach high precision(96.6$\%$) and recall(85.3$\%$) simultaneously, hence also a high F1 score (0.906). Further test of the trained PmPNet on the 2000.01 $\sim$ 2018.12 seismic data shows that the algorithm can successfully identify almost all PmP waves (larger than 96$\%$) before 2011, and still more than 85$\%$ of the PmP waves after 2010 where the test data are not involved in training the PmPNet. Applying the trained PmPNet to the seismic data during the period Jan. 1990.01 $\sim$ 1999.12, we have nearly tripled the volume of the PmP database in southern California (28,093 picks).}

The updated PmP database in this study provides valuable seismic observations that can help us image the whole crustal P-wave structure beneath southern California when complemented with the local first P-wave travel time data. Furthermore, the resulting high-resolution P-wave structure, together with the S-wave structure constrained by ambient noise data, will lay a solid foundation for our understanding of the various deformation process within the lower crust~\cite{shinevar2018inferring,B2008Rheology}. Besides, the high-quality PmP database in this study can serve as the template to help train the PmPNet and make it suitable for identifying the PmP wave from both the broadband and short-period seismic data. Moreover, we could build the PmP database in various tectonic settings following this way. Our developed PmPNet can be easily modified to recognize other later seismic phases, especially when they are relatively rarely used in routine seismic studies, because of the excellent performance of our designed algorithm on the unbalanced data.

\acknowledgments
The seismic data and local earthquake catalog are requested from Southern California Earthquake Data Center\footnote{\url{https://scedc.caltech.edu/}}. {The picked PmP waves and PmPNet code developed in this study can be accessed at GitHub\footnote{ \url{https://github.com/Seismic-Data-imaging-the-Earth/PmPWorld}}. The authors thank the Editor, Michael Bostock, and the Associate Editor for handling this paper. The authors are grateful to the Associate Editor, the reviewer Albert Leonardo and another anomalous reviewer for their constructive comments that have greatly improved the paper.} The work of W. Ding and K. Ren was partially supported by the NSF grants DMS-1937254 and EAR-2000850. T. Li and P. Tong were partly supported by Singapore MOE AcRF Tier-2 Grant (MOE2019-T2-2-112) and the National Research Foundation Singapore and the Singapore Ministry of Education under the Research Centers of Excellence Initiative (Project Code Number: 04MNS001953A620). X. Yang was partially supported by the NSF grants DMS-1818592 and DMS-2109116. Part of the work was initiated during X. Yang's visit to the Department of Applied Physics and Applied Mathematics, Columbia University, and X. Yang is grateful to the department for their hospitality. 

%Enter acknowledgments, including your data availability statement, here.

%% ------------------------------------------------------------------------ %%
%% References and Citations

%%%%%%%%%%%%%%%%%%%%%%%%%%%%%%%%%%%%%%%%%%%%%%%
%
% \bibliography{<name of your .bib file>} don't specify the file extension
%
% don't specify bibliographystyle
%%%%%%%%%%%%%%%%%%%%%%%%%%%%%%%%%%%%%%%%%%%%%%%

\bibliography{PmPref.bib}

%Reference citation instructions and examples:
%
% Please use ONLY \cite and \citeA for reference citations.
% \cite for parenthetical references
% ...as shown in recent studies (Simpson et al., 2019)
% \citeA for in-text citations
% ...Simpson et al. (2019) have shown...
%
%
%...as shown by \citeA{jskilby}.
%...as shown by \citeA{lewin76}, \citeA{carson86}, \citeA{bartoldy02}, and \citeA{rinaldi03}.
%...has been shown \cite{jskilbye}.
%...has been shown \cite{lewin76,carson86,bartoldy02,rinaldi03}.
%... \cite <i.e.>[]{lewin76,carson86,bartoldy02,rinaldi03}.
%...has been shown by \cite <e.g.,>[and others]{lewin76}.
%
% apacite uses < > for prenotes and [ ] for postnotes
% DO NOT use other cite commands (e.g., \citet, \citep, \citeyear, \nocite, \citealp, etc.).
%

\section*{Supporting Information for ``Deep Neural Networks for Creating Reliable PmP Database with a Case Study in Southern California''}
%
% e.g., \title{Supporting Information for "Terrestrial ring current:
% Origin, formation, and decay $\alpha\beta\Gamma\Delta$"}
%
%DOI: 10.1002/%insert paper number here%

%% ------------------------------------------------------------------------ %%
%
%  AUTHORS AND AFFILIATIONS
%
%% ------------------------------------------------------------------------ %%

% List authors by first name or initial followed by last name and
% separated by commas. Use \affil{} to number affiliations, and
% \thanks{} for author notes.
% Additional author notes should be indicated with \thanks{} (for
% example, for current addresses).

% Example: \authors{A. B. Author\affil{1}\thanks{Current address, Antartica}, B. C. Author\affil{2,3}, and D. E.
% Author\affil{3,4}\thanks{Also funded by Monsanto.}}

%\authors{=Authors=}

%% ------------------------------------------------------------------------ %%
%
%  BEGIN ARTICLE
%
%% ------------------------------------------------------------------------ %%

% The body of the article must start with a \begin{article} command
%
% \end{article} must follow the references section, before the figures
%  and tables.

%% ------------------------------------------------------------------------ %%
%
%  TEXT
%
%% ------------------------------------------------------------------------ %%

\begin{figure}
	\centering
	\includegraphics[width=0.9\linewidth]{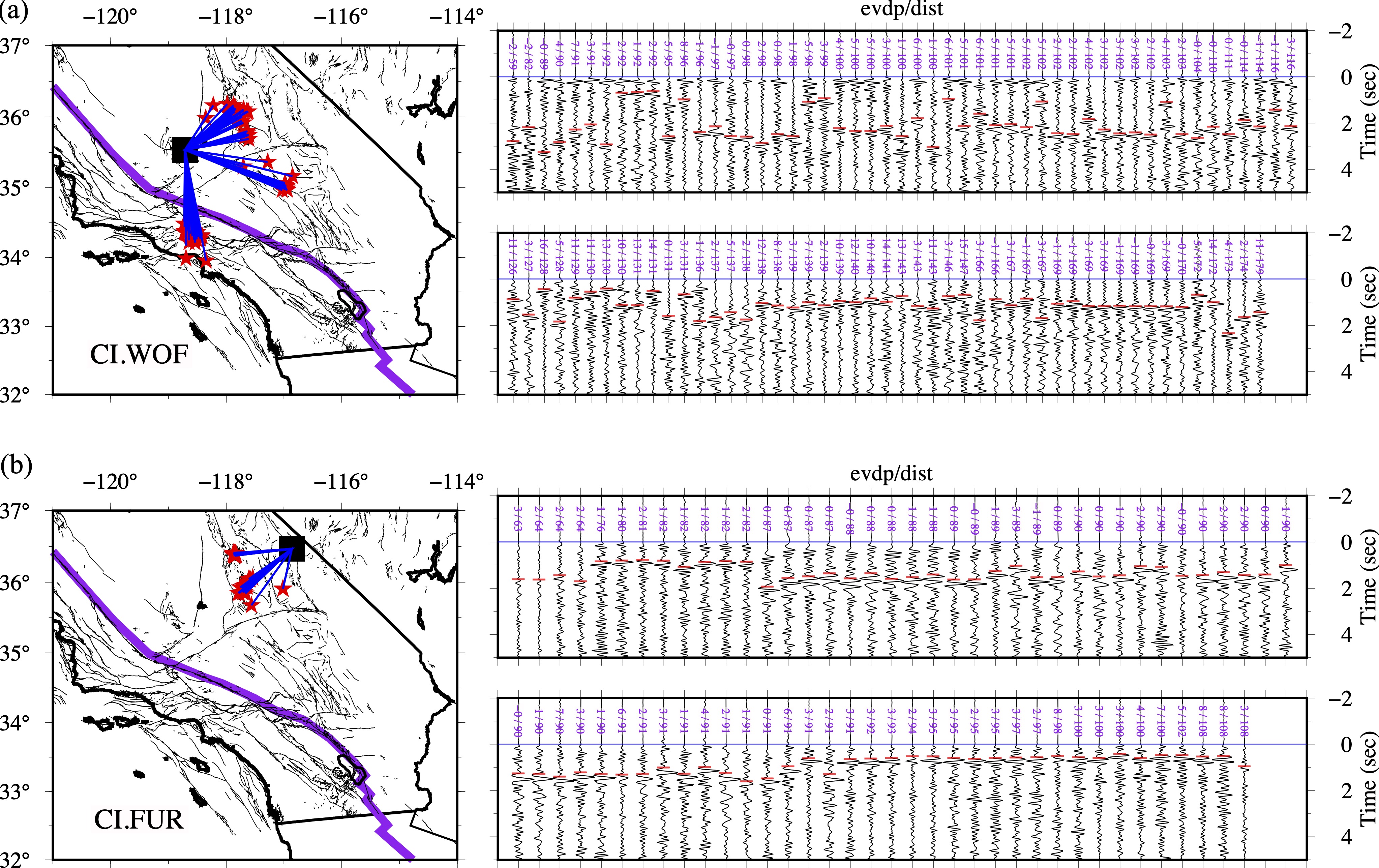}
	\caption{Example of the PmP waves recorded at the short-period station CI.WOF (a) and broadband station CI.FUR (b). In the map, the seismic station is denoted as the black-filled square, and local earthquakes are represented as the red-filled stars. For the PmP waveforms, they have been aligned to the first P-wave arrivals and normalized in the displaying window. The onset time for the PmP wave is indicated by the red bar. Numbers in purple show the earthquake depth (evdp) and the event-station distance (dist) for each waveform.}
	\label{FIG:si_fig1}
\end{figure}

\begin{figure}
	\centering
	\includegraphics[width=0.9\linewidth]{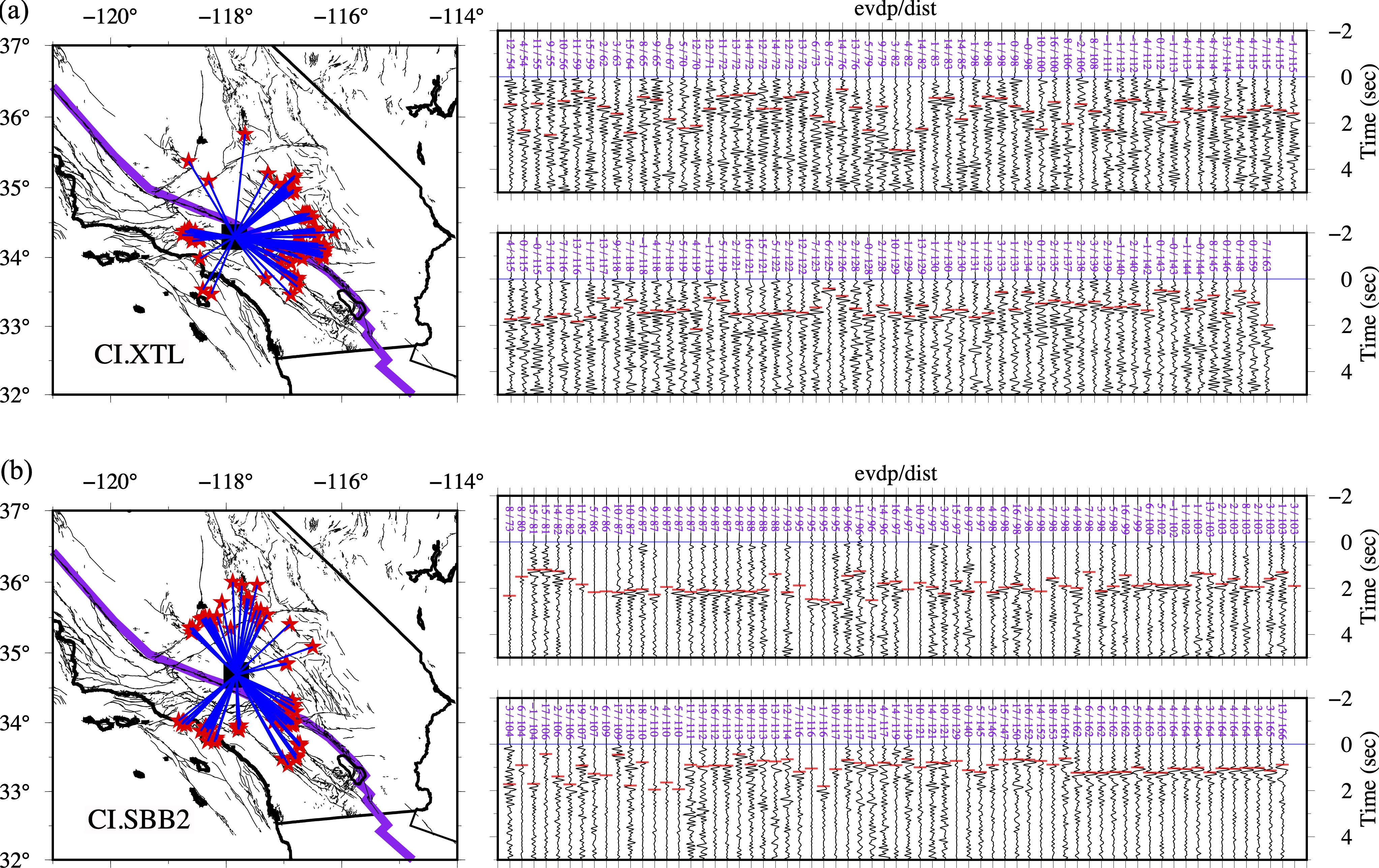}
	\caption{Example of the PmP waves recorded at the short-period station CI.XTL (a) and broadband station CI.SBB2 (b). See the detailed captions in Figure~\ref{FIG:si_fig1}.}
	\label{FIG:si_fig2}
\end{figure}

\begin{figure}
	\centering
	\includegraphics[width=0.9\linewidth]{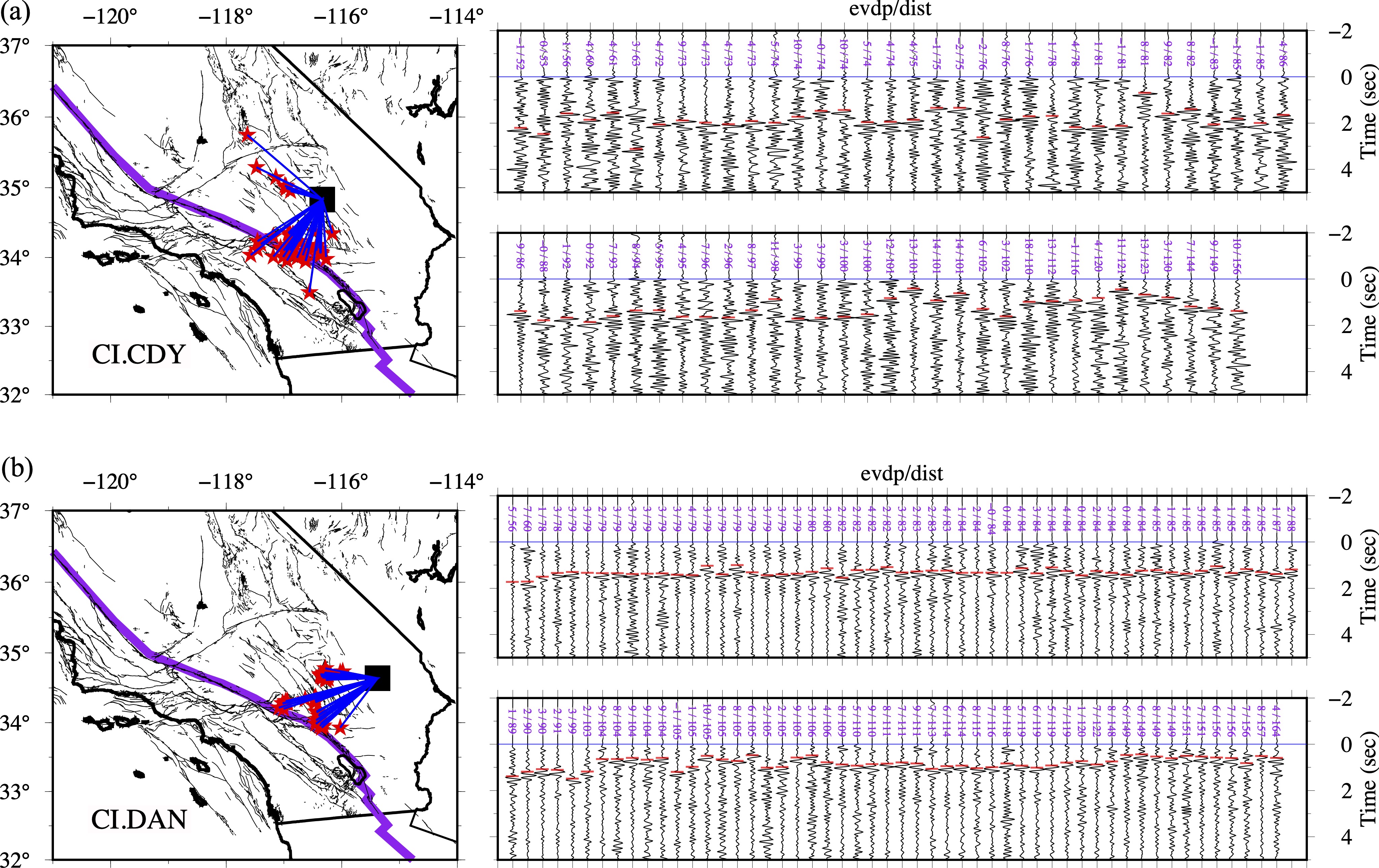}
	\caption{Example of the PmP waves recorded at the short-period station CI.CDY (a) and broadband station CI.DAN (b). See the detailed captions in Figure~\ref{FIG:si_fig1}.}
	\label{FIG:si_fig3}
\end{figure}

\begin{figure}
	\centering
	\includegraphics[width=0.9\linewidth]{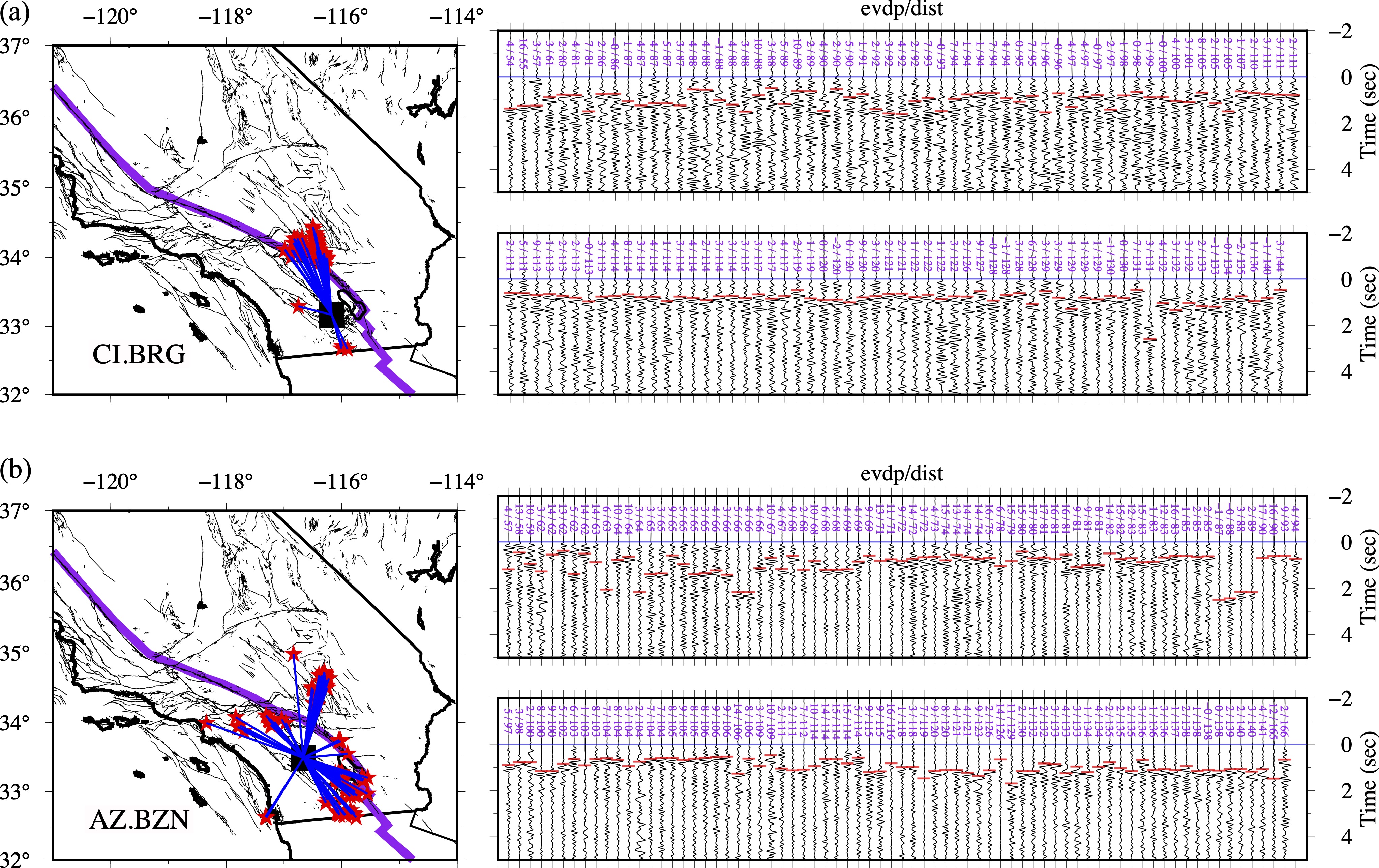}
	\caption{Example of the PmP waves recorded at the short-period station CI.BRG (a) and broadband station AZ.BZN (b). See the detailed captions in Figure~\ref{FIG:si_fig1}.}
	\label{FIG:si_fig4}
\end{figure}

\noindent\textbf{Contents of this file}
%%%Remove or add items as needed%%%
\begin{enumerate}
\item {Text S1}
\item Figures~~\ref{FIG:si_fig1} to ~\ref{FIG:si_fig7}
\item Table~\ref{tab:si_tab1}
%if Tables are larger than 1 page, upload as separate excel file
\end{enumerate}
%\noindent\textbf{Additional Supporting Information (Files %uploaded separately)}
%\begin{enumerate}
%\item Captions for Datasets S1 to Sx
%\item Captions for large Tables S1 to Sx (if larger than 1 %page, upload as separate excel file)
%\item Captions for Movies S1 to Sx
%\item Captions for Audio S1 to Sx
%\end{enumerate}

%\noindent\textbf{Introduction}
%Type or paste your text here. The introduction gives a brief overview of the supporting information. You should include information %about as many of the following as possible (when appropriate):
% 1. a general overview of the kind of data files;
% 2. information about when and how the data were collected or created;
% 3. a general description of processing steps used;
% 4. any known imperfections or anomalies in the data.
\noindent\textbf{Introduction}
These supporting materials include {the detailed introduction of PmP-traveltime-Net which is designed to improve the prediction accuracy of PmP traveltime (Text S1),} examples of both the short-period and broadband waveforms included in our newly built PmP database (Figures~\ref{FIG:si_fig1}-\ref{FIG:si_fig4}), tests of using different hyperparameters in training the PmPNet (Figure~\ref{FIG:si_fig5}), the comparison between the original waveforms and their associated spectrums, which are either recorded by a short-period seismic station or by a broadband seismic station (Figure~\ref{FIG:si_fig6}), {test performance of PmP-traveltime-Net (Figure~\ref{FIG:si_fig7})} and the statistical rareness of high-quality PmP waves in available seismic records (Table~\ref{tab:si_tab1}).

\clearpage

%Delete all unused file types below. Copy/paste for multiples of each file type as needed.
\noindent\textbf{{Text S1.}}
{To improve the prediction accuracy of PmP traveltime, we further design a separate model called “PmP-traveltime-Net” to predict the difference in traveltime between PmP and first P waves. The structure of PmP-traveltime-Net is very similar to PmPNet, except for the absolute PmP traveltime in output has been changed into the difference in traveltime between PmP and first P waves, and PmP-traveltime-Net only outputs one quantity: the predicted differential traveltime, which is a positive real number. We then compute the absolute PmP traveltime from source to receiver by adding the P-wave traveltime to the predicted differential traveltime. The loss function for PmP-traveltime-Net is defined as the absolute difference($L^1$ loss) between the true and predicted traveltimes. We use PmP-traveltime-Net only to predict the PmP traveltimes for the waveforms which are identified already to contain clear PmP waves. We train PmP-traveltime-Net on 10,000 waveform data with clear PmP waves prepared by the two-stage workflow \cite{li2022moho} and validate it on all the manually picked data (during the period from 1990 to 2018). Because of the efficiency of training PmP-traveltime-Net, it takes only $\sim$ 7.25 minutes to finish the 200-epochs training on one NVIDIA GeForce RTX 2080 graphics processing units (GPUs), and the training loss stabilizes at around 150th epoch (Figure~\ref{FIG:si_fig7}a). Test performance of PmP-traveltime-Net shows that there’re 95$\%$ of picks with traveltime residuals no larger than 0.5 s and even 88$\%$ of picks with traveltime residuals no larger than 0.1 s (Figures~\ref{FIG:si_fig7}b.)}

\clearpage

\begin{figure}
	\centering
	\includegraphics[width=0.9\linewidth]{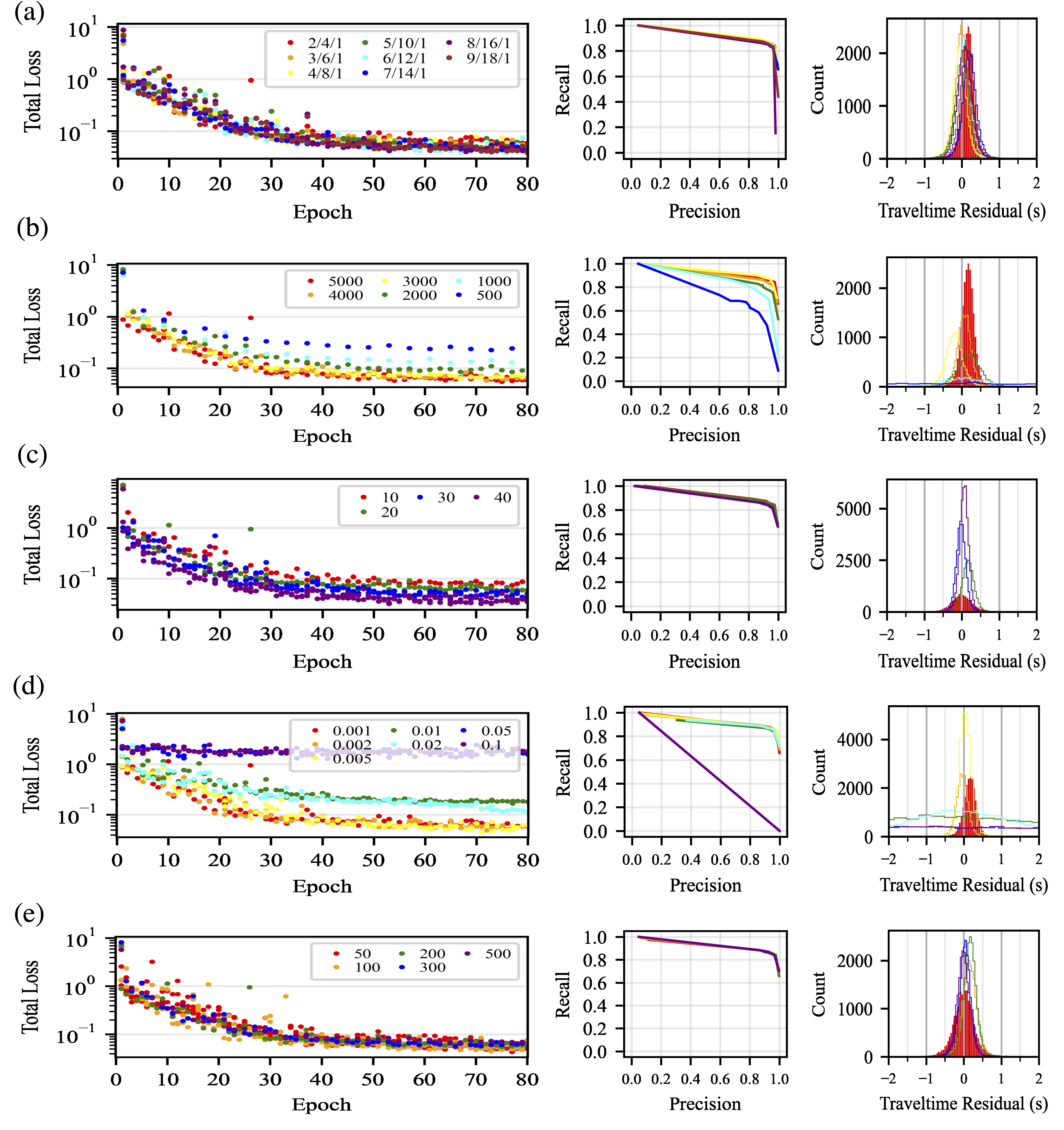}
%	\label{FIG:si_fig5}
\end{figure}
\begin{figure}
	\centering
	\caption{Tests for using different hyperparameters in training PmPNet. We show the test result for each group hyperparameters in 3 panels$:$ The total loss with respect to the epoch count, the recall versus precision curve, and histogram of traveltime residual between the picked/calculated PmP traveltime and that predicted by PmPNet. Results are color-coded with respect to the used hyperparameter in training PmPNet. (a) Number of ResNet blocks, with other hyperparameters fixed, e.g., learning rate $=$ 0.001, PmP amount $=$ 5000, rate(PmP/NonPmP) $=$ 1/20, batch size $=$ 200 and epoch number $=$ 80. (b) Number of the PmP-labeled data in the training dataset, with other hyperparameters fixed, e.g., learning rate $=$ 0.001, grid size $=$ 2/4/1, rate(PmP/NonPmP) $=$ 1/20, batch size $=$ 200 and epoch number $=$ 80. (c) The Non-PmP/PmP data amount ratio in the training dataset, with other hyperparameters fixed, e.g., learning rate $=$ 0.001, grid size $=$ 2/4/1, PmP amount $=$ 5,000, batch size $=$ 200 and epoch number $=$ 80. (d) Learning rate, with other hyperparameters fixed, e.g., grid size $=$ 2/4/1, PmP amount $=$ 5,000, rate(PmP/NonPmP) $=$ 1/20, batch size $=$ 200 and epoch number $=$ 80. (e) Batch size, with other hyperparameters fixed, e.g., learning rate $=$ 0.001, grid size $=$ 2/4/1, PmP amount $=$ 5,000, rate(PmP/NonPmP) $=$ 1/20 and epoch number $=$ 80.}
	\label{FIG:si_fig5}
\end{figure}

\begin{figure}
	\centering
	\includegraphics[width=0.9\linewidth]{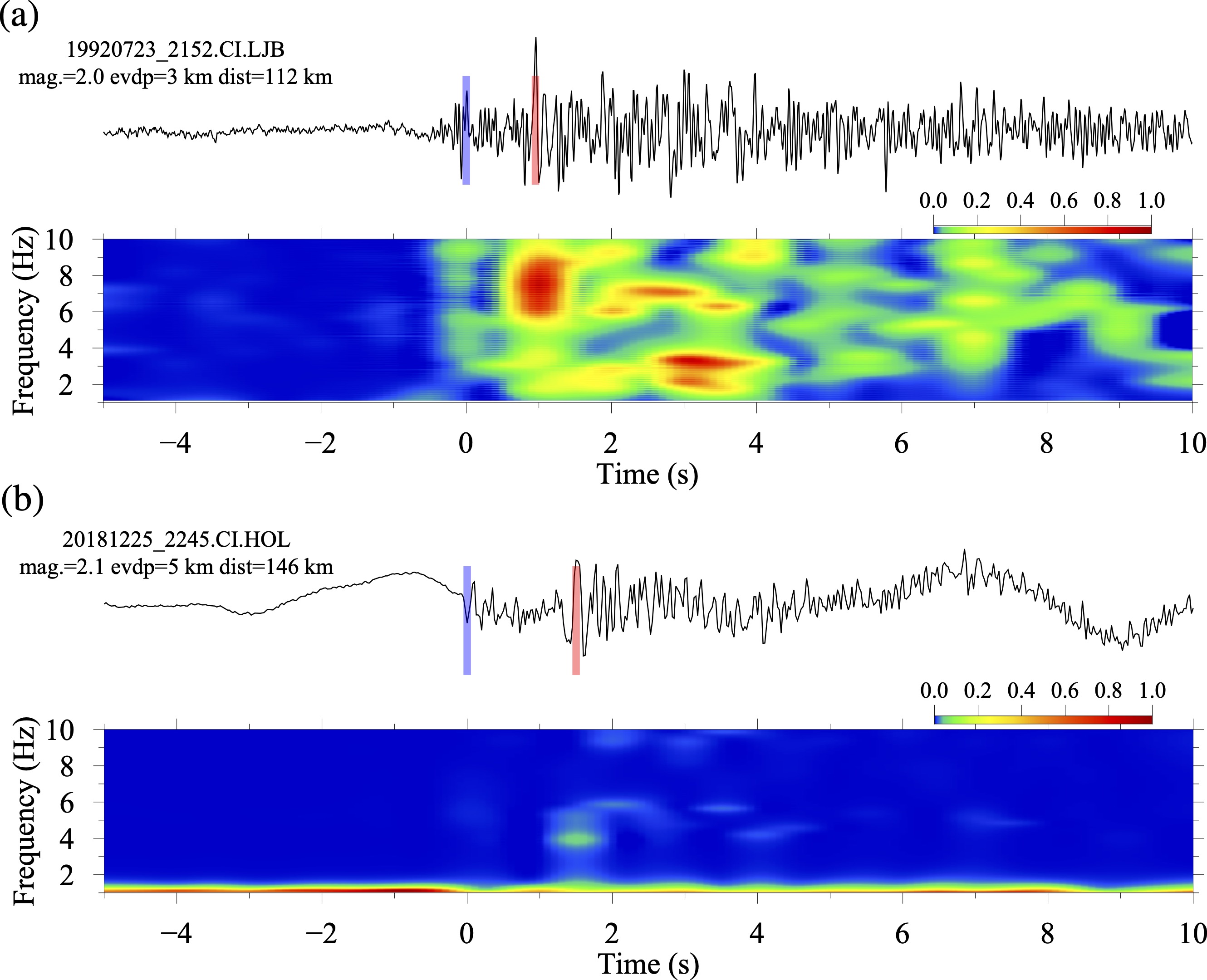}
	\caption{(a) An example of the original short-period waveform and its spectrum. The waveform is recorded at station CI.LJB and it is triggered by an earthquake with the magnitude of 2.0, focal depth of 3 km and epicentral distance of 112 km. Blue bar indicates the onset time of first P wave and red bar indicates the onset time of  PmP wave. (b) An example of the original broadband waveform and its spectrum. The waveform is recorded at station CI.HOL and it is triggered by an earthquake with the magnitude of 2.1, focal depth of 5 km and epicentral distance of 146 km.}
	\label{FIG:si_fig6}
\end{figure}

\begin{figure}
	\centering
	\includegraphics[width=0.9\linewidth]{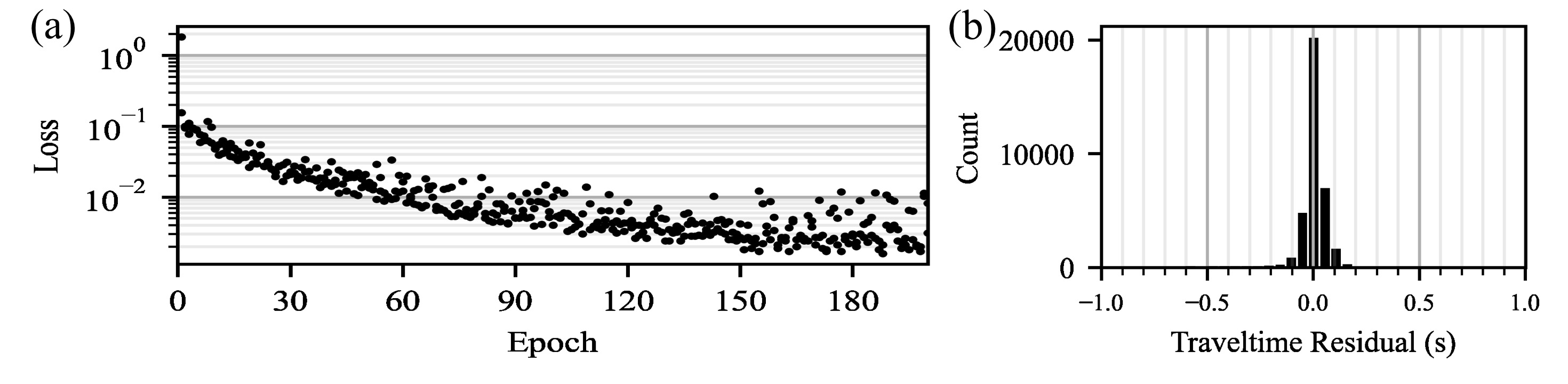}
	\caption{Test performance of PmP-traveltime-Net. We train PmP-traveltime-Net on 10,000 waveform data with clear PmP waves prepared by two-stage workflow  and validate it on all the manually picked data (during the period from 1990 to 2018). (a) The loss with respect to the epoch count. (b) Histogram of traveltime residual between the predicted PmP traveltime by PmP-traveltime-Net and those picked manually.}
	\label{FIG:si_fig7}
\end{figure}

\clearpage

\begin{table}[H]
\centering
%\settablenum{S1} %%Change number for each table
\caption{Rare high-quality PmP waveforms in available seismic records.}\label{tab:si_tab1}
\begin{tabular}{*{5}{c}}
\hline
 \thead{Study region} & \thead{Southern California \\ \cite{li2022moho}} & \thead{West Japan \\ \cite{wang2018crustal}} & \thead{Southwest Japan \\ \cite{sun2008seismic}} & \thead{Central Tohoku, Japan \\ \cite{xia2007mapping}}  \\
\hline
 Record period & 01/2000$\sim$12/2018 & 04/2004$\sim$03/2017 & 09/2002$\sim$09/2006 & 07/2002$\sim$10/2006 \\
 Record amount & 403,371 & 125,836 & 1,740 & 6,450 \\
 PmP amount & 10,192 & 2,235 & 478 & 394 \\
 \thead{Amount ratio \\ (PmP$/$Record)} & 2.5$\%$ & 1.8$\%$ & 27.5$\%$ & 6.1$\%$ \\
\hline
\end{tabular}
\end{table}

%\bibliography{PmPref.bib}

\end{document}